\documentclass[prd,twocolumn,reprint,superscriptaddress]{revtex4-2}
\usepackage{feynmp,amsmath,mathtools,amssymb,graphicx,subfigure,hyperref,bm,mathrsfs,slashed}
\usepackage{CJK}
\usepackage{color}

\begin{document}
\begin{CJK}{UTF8}{song}
		
\title{Reduction for one-loop tensor Feynman integrals in the relativistic quantum field theories at finite temperature and/or finite density}

\author{Hao-Ran Chang}
\thanks{hrchang@mail.ustc.edu.cn}
\affiliation{Department of Physics, Sichuan Normal University, Chengdu, Sichuan 610066, China}
\affiliation{Center for Computational Sciences, Sichuan Normal University, Chengdu, Sichuan 610066, China}
\affiliation{Department of Physics, McGill University, Montreal, Quebec H3A 2T8, Canada}

\date{\today}
		
\begin{abstract}
The \emph{conventional} Passarino-Veltman reduction is a systematic procedure based on the Lorentz covariance,
which can efficiently reduce the one-loop tensor Feynman integrals in the relativistic quantum field theories (QFTs)
at zero temperature and zero density. However, the Lorentz covariance is explicitly broken when either of the temperature and density is finite, due to a rest reference frame of the many-body system in which the temperature and density are
measured, rendering the \emph{conventional} Passarino-Veltman reduction not applicable anymore to reduce the one-loop tensor Feynman integrals therein. In this paper, we report a \emph{generalized} Passarino-Veltman reduction which can efficiently simplify the one-loop tensor Feynman integrals in the relativistic QFTs at finite temperature and/or finite density. The \emph{generalized} Passarino-Veltman reduction can analyze the one-loop tensor Feynman integrals in a wide range of physical systems described by the relativistic QFTs at finite temperature and/or finite density, such as quark-gluon plasma in nuclear physics.
\end{abstract}

\maketitle


\section{Introduction}

Lorentz symmetry is always respected by quantum fields with relativistic energy-momentum relation in vacuum where the temperature and density are both zero \cite{Weinberg}. However, this symmetry is explicitly broken when either of the temperature and density is finite, because a rest inertial reference frame of the many-body system is specified in which the temperature and density are measured \cite{Gale,Das,Nair,Bellac,Ezawa}. Consequently, in the relativistic quantum field theories (QFTs) thereat, the Lorentz symmetry is violated, which definitely dictates a series of substantial physical consequences encoded into the corresponding transition amplitudes and correlation functions. 

One-loop Feynman diagrams play an extremely important role in calculating the transition amplitudes and correlation functions in many subfields of physics, including particle physics, nuclear physics, and condensed matter physics \cite{Weinberg,Gale,Das,Nair,Bellac,Ezawa,Fetter,Abrikosov,Mahan}. Unfortunately, the one-by-one evaluation of one-loop Feynman diagrams is often rather cumbersome, time-consuming, error-prone, and disposable, given the amount of one-loop Feynman diagrams that needs to be computed. Therefore, a very significant problem is how to most efficiently perform the calculation of one-loop Feynman diagrams. 

Pioneering works in calculating the one-loop Feynman diagrams for the relativistic QFTs at zero temperature and zero density were performed by Veltman and his collaborators. Elaborating on the original idea of Brown and Feynman \cite{Feynman}, Passarino and Veltman provided a systematic reduction scheme of generic one-loop tensor Feynman integrals based on the Lorentz covariance. With the help of \emph{conventional} Passarino-Veltman reduction\cite{CPVRS}, the generic one-loop tensor Feynman integrals can be reduced to a small number of generic one-loop scalar Feynman integrals \cite{CPVRS}. As a result, the calculations of a huge number of one-loop Feynman diagrams boil down to nothing but the automatic assembly of a small number of generic one-loop scalar Feynman integrals \cite{TVSI,CPVRS,Ellis,Denner2006}. Specifically, the generic one-loop scalar Feynman integrals up to four-point have been analytically calculated by `t Hooft and Veltman \cite{TVSI}, and their analytical properties have also been well studied \cite{Karplus1958,Frederiksen1971,Rechenberg1972,Frederiksen1973,Denner1993}. Acting as the basic building blocks, the generic one-loop scalar Feynman integrals are reusable to the calculation of one-loop Feynman diagrams in various physical processes and different relativistic quantum fields. Owing to this advantage, the \emph{conventional} Passarino-Veltman reduction triggered many variants \cite{Denner2006,Ezawa1992,Tarasov1996,Aguila2004,Hameren2005,
Beanger2006,Ellis2006,Denner2003,Battistel2006,HRC2012,Battistel2012,Oldenborgh1990}, and the according program packages for automatic algebraic calculation \cite{Oldenborgh1990,FeynArts1990,FeynCalc1991,LoopTools1999,QCDLoop2008,PackageX2015} help to systematize the calculation of one-loop Feynman diagrams and have been widely applied in numerous processes of high energy physics.

The precondition of applying \emph{conventional} Passarino-Veltman reduction \cite{CPVRS} and its variants \cite{Ellis,Denner2006,Ezawa1992,Tarasov1996,Aguila2004,
Hameren2005,Beanger2006,Ellis2006,Denner2003,Battistel2006,HRC2012,Battistel2012,Oldenborgh1990} is that the physical systems
of interest must respect the Lorentz symmetry, which is satisfied only for the relativistic QFTs at zero temperature and zero density. When either the temperature or the density is finite, the \emph{conventional} Passarino-Veltman reduction is not applicable any more to reduce the generic one-loop tensor Feynman integrals due to the explicit breaking of Lorentz symmetry therein. A natural question follows that what the counterpart of \emph{conventional} Passarino-Veltman reduction is in the relativistic QFTs when either of the temperature and density is finite. On the other hand, it is known that the calculation of one-loop Feynman diagrams in the relativistic QFTs become more complicated when either the temperature or the density is not zero any more \cite{Weinberg,Gale,Das,Nair,Bellac,Ezawa,Fetter,Abrikosov,Mahan}. Unfortunately, there have only been sparse attempts \cite{Rehberg1996PRC, Rehberg1996AOP,HRC2018} aiming at generalizing the \emph{conventional} Passarino-Veltman reduction to its counterpart at finite temperature and/or finite density. Although several generic one-loop scalar Feynman integrals up to three-point had been analytically calculated \cite{Rehberg1996AOP}, just as their counterparts at zero temperature and zero density calculated by `t Hooft and Veltman \cite{TVSI}, the reduction procedure for the generic one-loop tensor Feynman integrals still remains undeveloped, limiting the efficient application of these generic one-loop scalar Feynman integrals to various physical processes and different interactions. Hence, from both the theoretical interest and application-driven consideration, it is in need to develop a systematic reduction for efficiently evaluating the one-loop Feynman diagrams in the relativistic QFTs when either of the temperature and density is finite. 

Motivated by these, we report a \emph{generalized} Passarino-Veltman reduction in this work, which can simplify a huge amount of one-loop Feynman diagrams in a wide range of physical systems described by the relativistic QFTs at finite temperature and/or finite density, such as quark-gluon plasma in nuclear physics \cite{QGP1,QGP2}. The rest of this paper is organized as follows. In Sec.\ref{GOLSTIs}, we present the general definitions of $N$-point generic one-loop scalar Feynman integrals and one-loop tensor Feynman integrals in the relativistic QFTs at finite temperature and/or finite density, and emphasize the explicit breaking of Lorentz covariance due to either the finite temperature or finite density and its consequences in the reduction of one-loop Feynman diagrams. We show the detailed reduction procedures of one-point, two-point, and three-point generic one-loop tensor Feynman integrals in Sec. \ref{ROLTFI}. Two demonstration applications of the \emph{generalized} Passarino-Veltman reduction are shown in Sec. \ref{Applications}. Our main conclusions and discussions are presented in Sec. \ref{Summary}. Finally, we provide three appendices to show some detailed calculation.

\begin{widetext}

\section{Generic one-loop scalar and tensor Feynman integrals\label{GOLSTIs}}

A direct comparison among the one-loop Feynman diagrams in various physical processes and different theoretical models led to an observation that any one-loop Feynman diagrams in the relativistic QFTs can be decomposed into a linear combination of generic one-loop tensor Feynman integrals and generic one-loop scalar Feynman integrals \cite{CPVRS,Feynman,Ellis,Denner2006} (see Sec. \ref{Applications} for demonstrations).

\begin{figure}[htbp]
	\centering
	\includegraphics[width=7cm]{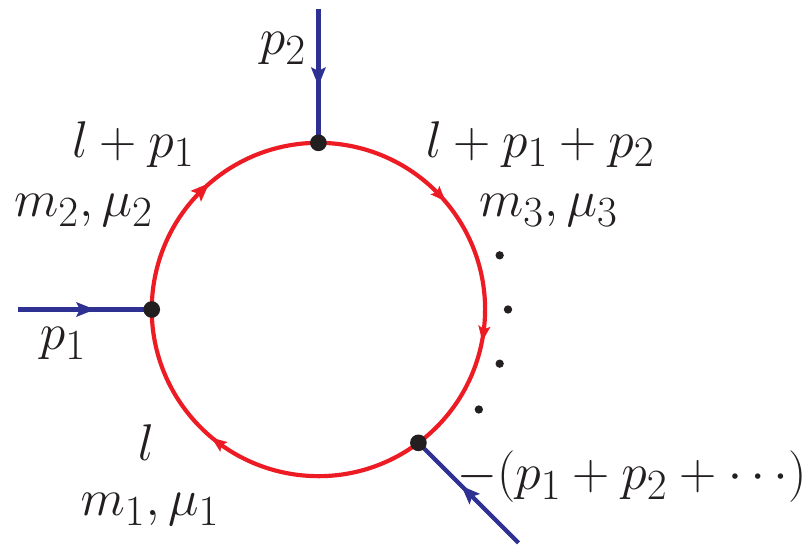}
	\caption{Generic $N$-point one-loop Feynman diagram with external momenta $p_n=(p_n^0,\boldsymbol{p}_n)=(p_n^0,p_n^i)$. The momentum in the first internal propagator is denoted by $l=(l^0,\boldsymbol{l})=(l^0,l^i)$. All Lorentz momenta are defined to be
		directed towards the vertices. In addition, $m_{n}$ and $\mu_{n}$ represent the mass and density (chemical potential) in the $\textit{n}$th propagator, respectively.}
	\label{FigNpts}
\end{figure}

 The $N$-point generic scalar Feynman integral and generic tensor Feynman integrals in the one-loop Feynman diagrams (see Fig. \ref{FigNpts}) have the general form

\begin{align}
\mathcal{I}_{N}(p_1,\cdots,p_{N-1};m_1,\mu_1;\cdots;m_N,\mu_N;\beta)
&=\int\frac{d^{D}l}{i\pi^{D/2}}\frac{1}
	{\mathcal{D}(l;p_1,\cdots,p_{N-1};m_1,\mu_1;\cdots;m_N,\mu_N)},\\
\mathscr{I}_{N}^{\left\{\rho;\rho\sigma;\rho\sigma\tau;\cdots\right\}}(p_1,\cdots,p_{N-1};m_1,\mu_1;\cdots;m_N,\mu_N;\beta)
&=\int\frac{d^{D}l}{i\pi^{D/2}}\frac{\left\{l^{\rho};l^{\rho}l^{\sigma};l^{\rho}l^{\sigma}l^{\tau};\cdots\right\}}
	{\mathcal{D}(l;p_1,\cdots,p_{N-1};m_1,\mu_1;\cdots;m_N,\mu_N)},
\end{align}
where $\beta=1/(k_B T)$ with $k_B$ the Boltzmann constant and $T$ the temperature, and the denominator of integrand is defined as
\begin{align}
\hspace{-0.25cm}
\mathcal{D}_{N}(l;p_1,\cdots,p_{N-1};m_1,\mu_1;\cdots;m_N,\mu_N)&=
\mathcal{P}(l,0;m_1,\mu_1)\mathcal{P}(l,p_1;m_2,\mu_2)\cdots\mathcal{P}(l,p_1+p_2+\cdots+p_{N-1};m_N,\mu_N),
\end{align}
with 
\begin{align}
\mathcal{P}(l,p;m_n,\mu_n)
&=(l^0+p^0+\mu_n)^2-\left[(\boldsymbol{l}+\boldsymbol{p})^2+m_n^2\right].
\end{align} 
The numerator of integrand in the generic one-loop scalar Feynman integral (rank-zero one-loop tensor Feynman integrals) $\mathcal{I}_{0}(p_1,\cdots,p_{N-1};m_1,\mu_1;\cdots;m_N,\mu_N;\beta)$ is replaced by $l^{\rho}$ for the rank-one generic one-loop tensor Feynman integral $\mathscr{I}^{\rho}(p_1,\cdots,p_{N-1};m_1,\mu_1;\cdots;m_N,\mu_N;\beta)$, by $l^{\rho}l^{\sigma}$ for the rank-two generic one-loop tensor Feynman integral $\mathscr{I}^{\rho\sigma}(p_1,\cdots,p_{N-1};m_1,\mu_1;\cdots;m_N,\mu_N;\beta)$, by $l^{\rho}l^{\sigma}l^{\tau}$ for the rank-three generic one-loop tensor Feynman integral $\mathscr{I}^{\rho\sigma\tau}(p_1,\cdots,p_{N-1};m_1,\mu_1;\cdots;m_N,\mu_N;\beta)$, and so forth.
In addition, $l=(l^0,\boldsymbol{l})=(l^0,l^i)$ is the momentum of the first internal propagator, $p_n=(p_n^0,\boldsymbol{p}_n)=(p_n^0,p_n^i)$ are the external momenta, $m_{n}$ and $\mu_{n}$ denote the mass and density (chemical potential) in the $\textit{n}$th internal propagators, respectively. All the Lorentz momenta are defined to be
directed towards the vertices. The temperature is encoded into the temporal component of internal momentum, the loop integration over which is needed to be replaced by the summation over the Matsubara frequency  \cite{Matsubara} only when the explicit analytical calculations of the generic one-loop scalar Feynman integrals and the purely temporal components of generic one-loop tensor Feynman integrals are involved. However, such explicit analytical calculations are beyond the scope of our present work. Throughout this paper, we preserve the integration over $l^{0}$ instead of the summation over the Matsubara frequency, and hence the temporal component of internal momentum is a continuous variable allowing an arbitrary shift. Consequently, the temperature does not explicitly influence the reduction procedure of generic one-loop tensor Feynman integrals. In addition, we label the spacetime (spatial) components by Greek (Roman) letters, work in Minkowski space with a metric, $g_{\rho\sigma}=g^{\rho\sigma}=\mathrm{diag}(1,-1,\cdots,-1)$, and take the natural units where $\hbar=c=1$.

The one-loop Feynman integrals in the relativistic QFTs can be classified by $N$ (the number of external momenta) and $r$ (the rank of tensors of loop momentum in the numerator of integrand) \cite{Ellis,Denner1993}. Based on power counting from the Feynman rules in the relativistic QFTs, one observes that the coupling constants can regulate the rank of tensors of the loop momenta in the numerator of integrand in arbitrary loop Feynman diagrams therein. In the $N$-point one-loop Feynman diagrams of any renormalizable relativistic QFTs (such as the Standard Model and quantum chromodynamics) or super-renormalizable relativistic QFTs, the rank of tensors of the loop momentum turn out to be suppressed by the coupling constants of non-negative mass dimensions. Consequently, the highest rank ($r_h$) of tensor of the loop momentum in the integrand does not exceed the number ($N$) of external momenta, namely, $r_h \le N$. This is the reason that one usually stops at a given rank $r_h=N$ and does not need to proceed further to arbitrarily high ranks \cite{Ellis,Denner1993}. In the $N$-point one-loop Feynman diagrams of any nonrenormalizable relativistic QFTs, the coupling constants of negative mass dimension therein tend to increase the rank of tensors of the loop momentum. As a consequence, the highest rank ($r_h$) of tensors of the loop momentum in the numerator of integrand can exceed the number of external momenta, namely, $r_h\ge N$. With the procedure proposed in this paper, the reduction for tensor Feynman integral can be trivially extended to higher ranks and hence applicable to the one-loop tensor Feynman integrals in nonrenormalizable theories.

Since the calculation of one-loop Feynman diagrams in the renormalizable relativistic QFTs, such as the Standard Model and quantum chromodynamics, is especially useful for studying the physical systems including quark-gluon plasma and nuclear/hadronic matter \cite{Gale,Das,Nair,Bellac,Ezawa}, we restrict our attention in this work to the tensor reduction of one-loop tensor Feynman integrals in renormalizable relativistic QFTs, where the rank of one-loop tensor Feynman integrals can be set to be no more than the number of external momenta, namely, $r_h\le N$.

For the sake of simplicity and without loss of generality, we thereafter focus on the one-loop Feynman diagrams up to three-point ($N=3$) for illustrating the essential spirit of \emph{generalized} Passarino-Veltman reduction, and the generalization to $N$-point one-loop Feynman diagrams is straightforward. Up to three-point, any one-loop Feynman diagrams in the renormalizable relativistic QFTs can be decomposed into linear combinations of a series of generic one-loop scalar Feynman integrals,
\begin{align}
	\mathcal{A}_{0}(p;m_1,\mu_1;\beta)&=\mathcal{I}_{1}(p;m_1,\mu_1;\beta)
    \nonumber\\&
    =\int\frac{d^{D}l}{i\pi^{D/2}}\frac{1}{\mathcal{P}(l,p;m_1,\mu_1)},\\
	\mathcal{B}_{0}(p_1;m_1,\mu_1;m_2,\mu_2;\beta)&=\mathcal{I}_{2}(p_1;m_1,\mu_1;m_2,\mu_2;\beta)
    \nonumber\\&
    =\int\frac{d^{D}l}{i\pi^{D/2}}\frac{1}
	{\mathcal{P}(l,0;m_1,\mu_1)\mathcal{P}(l,p_1;m_2,\mu_2)},\\
	\mathcal{C}_{0}(p_1,p_2;m_1,\mu_1;m_2,\mu_2;m_3,\mu_3;\beta)&=\mathcal{I}_{3}(p_1,p_2;m_1,\mu_1;m_2,\mu_2;m_3,\mu_3;\beta)
    \nonumber\\&
    =\int\frac{d^{D}l}{i\pi^{D/2}}\frac{1}
	{\mathcal{P}(l,0;m_1,\mu_1)\mathcal{P}(l,p_1;m_2,\mu_2)\mathcal{P}(l,p_1+p_2;m_3,\mu_3)},
\end{align}
and generic one-loop tensor Feynman integrals,
\begin{align}
	\mathscr{A}^{\rho}(p;m_1,\mu_1;\beta)&=\mathscr{I}_{1}^{\rho}(p;m_1,\mu_1;\beta)
    \nonumber\\&
    =\int\frac{d^{D}l}{i\pi^{D/2}}\frac{l^{\rho}}{\mathcal{P}(l,p;m_1,\mu_1)},\\
	\mathscr{B}^{\left\{\rho;\rho\sigma\right\}}(p_1;m_1,\mu_1;m_2,\mu_2;\beta)
&=\mathscr{I}_{2}^{\left\{\rho;\rho\sigma\right\}}(p_1;m_1,\mu_1;m_2,\mu_2;\beta)
	\nonumber\\&
    =\int\frac{d^{D}l}{i\pi^{D/2}}\frac{\left\{l^{\rho};l^{\rho}l^{\sigma}\right\}}
	{\mathcal{P}(l,0;m_1,\mu_1)\mathcal{P}(l,p_1;m_2,\mu_2)},\\
	\mathscr{C}^{\left\{\rho;\rho\sigma;\rho\sigma\tau\right\}}(p_1,p_2;m_1,\mu_1;m_2,\mu_2;m_3,\mu_3;\beta)
    &=\mathscr{I}_{3}^{\left\{\rho;\rho\sigma;\rho\sigma\tau\right\}}(p_1,p_2;m_1,\mu_1;m_2,\mu_2;m_3,\mu_3;\beta)
	\nonumber\\&
    =\int\frac{d^{D}l}{i\pi^{D/2}}\frac{\left\{l^{\rho};l^{\rho}l^{\sigma};l^{\rho}l^{\sigma}l^{\tau}\right\}}
	{\mathcal{P}(l,0;m_1,\mu_1)\mathcal{P}(l,p_1;m_2,\mu_2)\mathcal{P}(l,p_1+p_2;m_3,\mu_3)}.
\end{align}
The compact notations in the definitions are explained as follows. First, following the notation of Ref. \cite{TVSI}, we set the symbols of one-loop scalar Feynman integral $\mathcal{I}_{N}$ and one-loop tensor Feynman integral $\mathscr{I}_{N}$ to be $\mathcal{A}_{0}$ and $\mathscr{A}$ for $N=1$, $\mathcal{B}_{0}$ and $\mathscr{B}$ for $N=2$, and $\mathcal{C}_{0}$ and $\mathscr{C}$ for $N=3$, respectively. Second, for the one-point one-loop tensor Feynman integral $\mathscr{A}^{\rho}(p;m_1,\mu_1;\beta)$, the numerator of integrand in the one-point one-loop scalar Feynman integral $\mathcal{A}_{0}(p;m_1,\mu_1;\beta)$ is replaced by $l^{\rho}$. Similarly for the two-point one-loop tensor Feynman integrals $\mathscr{B}^{\rho}(p_1;m_1,\mu_1;m_2,\mu_2;\beta)$ and $\mathscr{B}^{\rho\sigma}(p_1;m_1,\mu_1;m_2,\mu_2;\beta)$, the numerator of integrand in the two-point one-loop scalar Feynman integral $\mathcal{B}_{0}(p_1;m_1,\mu_1;m_2,\mu_2;\beta)$ are replaced by $l^{\rho}$ and $l^{\rho}l^{\sigma}$, respectively. The convention also holds for the three-point one-loop tensor Feynman integrals $\mathscr{C}^{\rho}(p_1,p_2;m_1,\mu_1;m_2,\mu_2;m_3,\mu_3;\beta)$, $\mathscr{C}^{\rho\sigma}(p_1,p_2;m_1,\mu_1;m_2,\mu_2;m_3,\mu_3;\beta)$, and $\mathscr{C}^{\rho\sigma\tau}(p_1,p_2;m_1,\mu_1;m_2,\mu_2;m_3,\mu_3;\beta)$, in which the numerators of integrand in the three-point one-loop scalar Feynman integral $\mathcal{C}_{0}(p_1,p_2;m_1,\mu_1;m_2,\mu_2;m_3,\mu_3;\beta)$ are replaced by $l^{\rho}$, $l^{\rho}l^{\sigma}$, and $l^{\rho}l^{\sigma}l^{\tau}$, respectively. Third, the generic one-loop tensor Feynman integrals $\mathscr{B}^{\rho\sigma}(p_1;m_1,\mu_1;m_2,\mu_2;\beta)$ and $\mathscr{C}^{\rho\sigma}(p_1,p_2;m_1,\mu_1;m_2,\mu_2;m_3,\mu_3;\beta)$ are symmetric in the Lorentz tensor indices $\rho$ and $\sigma$, and $\mathscr{C}^{\rho\sigma\tau}(p_1,p_2;m_1,\mu_1;m_2,\mu_2;m_3,\mu_3;\beta)$ is symmetric in the Lorentz tensor indices $\rho$, $\sigma$, and $\tau$.

Before going further, we emphasize the explicit breaking of Lorentz covariance due to finite temperature and/or finite density in the relativistic QFTs and its consequences. In the vacuum where the temperature and density are both zero, there is no preferred frame of reference. Therefore, the Lorentz covariance always holds for the relativistic QFTs, and consequently forces the amplitudes of one-loop Feynman diagrams therein (and hence the generic one-loop scalar Feynman integrals and one-loop tensor Feynman integrals) to be convariant functions of external momenta. Accordingly, with the help of \emph{conventional} Passarino-Veltman reduction, the generic one-loop tensor Feynman integrals can be reduced to the generic one-loop scalar Feynman integrals. However, in the presence of matter, the finite temperature and/or finite density select out a preferred rest frame of the heat bath, in which the temperature and density of the equilibrium thermal system are measured. The choice of this specific Lorentz frame of the many-body system explicitly breaks the Lorentz covariance \cite{Gale,Das,Nair,Bellac,Ezawa}. Consequently, the largest continuous spacetime symmetry of relativistic QFTs in $D$ dimension is no longer the $\mathrm{SO}(1,D-1)$ (proper normal) Lorentz symmetry but the $\mathrm{SO}(D-1)$ spatial rotation symmetry. Therefore, the expressions of one-loop Feynman diagrams (and hence the generic one-loop scalar Feynman integrals and one-loop tensor Feynman integrals) depend independently on the temporal component and the spatial component of external momenta. This leads to two crucial consequences regarding the incompleteness in both tensor structures and generic one-loop scalar Feynman integrals for reducing the generic one-loop tensor Feynman integrals in the relativistic QFTs at finite temperature and/or finite density. Firstly, the Lorentz-covariant tensor structures in the \emph{conventional} Passarino-Veltman reduction are no longer complete to expand the generic one-loop tensor Feynman integrals which are not Lorentz covariant. Secondly,
the generic one-loop scalar Feynman integrals $\mathcal{A}_{0}(p;m_1,\mu_1;\beta)$, $\mathcal{B}_{0}(p_1;m_1,\mu_1;m_2,\mu_2;\beta)$, and $\mathcal{C}_{0}(p_1,p_2;m_1,\mu_1;m_2,\mu_2;m_3,\mu_3;\beta)$ in the \emph{conventional} Passarino-Veltman reduction are incomplete to expand all the generic one-loop tensor Feynman integrals due to the explicit breaking of Lorentz covariance. It is stressed that these two aspects are the essential differences between the \emph{conventional} Passarino-Veltman reduction and \emph{generalized} Passarino-Veltman reduction, which is also the very motivation for the author to develop the \emph{generalized} Passarino-Veltman reduction.

The first incompleteness can be complemented by introducing an extra $D$-dimensional constant vector $u_{\rho}=u^{\rho}=(1;0,\cdots,0)$ in energy-momentum space, which was widely adopted to characterize the absence of Lorentz covariance due to the finite temperature and/or finite density \cite{Gale,Das,Nair,Bellac,Ezawa}. It is emphasized that this $D$-dimensional constant vector is defined only at finite temperature and/or finite density. By contrast, at zero temperature and zero density, this $D$-dimensional constant vector cannot play any role and is not defined due to the lacking of preferred rest frame \cite{Gale}. Treating the spacetime components of the generic one-loop tensor Feynman integrals on the same footing \cite{CPVRS,Ellis,Denner2006} and adding the effect of Lorentz-covariance breaking in terms of the constant vector, the generic one-loop tensor Feynman integrals can be reduced in the following section. For the second incompleteness, two generic two-point one-loop tensor Feynman integrals $\mathscr{B}^{0}(p_1;m_1,\mu_1;m_2,\mu_2;\beta)$ and $\mathscr{B}^{00}(p_1;m_1,\mu_1;m_2,\mu_2;\beta)$ and three generic three-point one-loop tensor Feynman integrals $\mathscr{C}^{0}(p_1,p_2;m_1,\mu_1;m_2,\mu_2;m_3,\mu_3;\beta)$, $\mathscr{C}^{00}(p_1,p_2;m_1,\mu_1;m_2,\mu_2;m_3,\mu_3;\beta)$, and $\mathscr{C}^{000}(p_1,p_2;m_1,\mu_1;m_2,\mu_2;m_3,\mu_3;\beta)$ must be introduced to form a complete set of generic one-loop Feynman integrals as the building blocks to express the form factors of the corresponding tensor structures. These two observations motivate us to develop the \emph{generalized} Passarino-Veltman reduction in this work for going beyond the applicability of previous works \cite{CPVRS,Rehberg1996PRC, Rehberg1996AOP,HRC2018}.

\section{Reduction of generic one-loop tensor Feynman integrals\label{ROLTFI}}

The essential spirit of \emph{conventional} Passarino-Veltman reduction is to decompose the generic one-loop tensor Feynman integrals into the corresponding generic one-loop scalar Feynman integrals with the help of Lorentz-covariant tensor structures and Lorentz-invariant generic one-loop scalar Feynman integrals. Different from that for the \emph{conventional} Passarino-Veltman reduction, the tensor structures for the \emph{generalized} Passarino-Veltman reduction contain the Lorentz-covariant parts and non-Lorentz-covariant parts and the generic one-loop tensor Feynman integrals in the \emph{generalized} Passarino-Veltman reduction must be decomposed into the corresponding generic one-loop scalar Feynman integrals and several purely temporal components of generic one-loop tensor Feynman integrals. In the following, we present the detailed reduction of one-point, two-point, and three-point generic one-loop tensor Feynman integrals, respectively.

\subsection{Reduction of one-point generic one-loop tensor Feynman integral $\mathscr{A}^{\rho}(p;m_1,\mu_1;\beta)$}

Before reducing the one-point generic one-loop tensor Feynman integral $\mathscr{A}^{\rho}(p;m_1,\mu_1;\beta)$, it is interesting to note that the one-point generic one-loop scalar Feynman integral,
\begin{align}
\mathcal{A}_{0}(p;m_1,\mu_1;\beta)&\equiv \int\frac{d^{D}l}{i\pi^{D/2}}\frac{1}{\mathcal{P}(l,p;m_1,\mu_1)}
=\int\frac{d^{D}l}{i\pi^{D/2}}\frac{1}{\mathcal{P}(l,0;m_1,\mu_1)}
\nonumber\\&
=\mathcal{A}_{0}(0;m_1,\mu_1;\beta)
\end{align}
by shifting the loop momentum $l$ to $(l-p)$, which indicates that $\mathcal{A}_{0}(p;m_1,\mu_1;\beta)$, is a function independent of external momentum $p$. It is emphasized that the temperature is encoded into the temporal component of internal momentum, the loop integration over which is needed to be replaced by the summation over the Matsubara frequency only when the explicit analytical calculations of the generic one-loop scalar Feynman integrals and the purely temporal components of generic one-loop tensor Feynman integrals are involved. However, such explicit analytical calculations are beyond the scope of our present work. Throughout this paper, we preserve the integration over $l^{0}$ instead of the summation over the Matsubara frequency, and hence the temporal component of internal momentum is a continuous variable allowing an arbitrary shift. 

The temporal component and spatial component of one-point generic one-loop tensor Feynman integral $\mathscr{A}^{\rho}(p;m_1,\mu_1;\beta)$ read
\begin{align}
\mathscr{A}^{0}(p;m_1,\mu_1;\beta)&=\int\frac{d^{D}l}{i\pi^{D/2}}\frac{(l^0+p^0+\mu_1)-(p^0+\mu_1)}{\mathcal{P}(l,p;m_1,\mu_1)},\\
\mathscr{A}^{i}(p;m_1,\mu_1;\beta)&\equiv \int\frac{d^{D}l}{i\pi^{D/2}}\frac{(l^i+p^i)-p^i}{\mathcal{P}(l,p;m_1,\mu_1)}.
\end{align}
After taking the asymmetry of integrands over symmetric domain of integrals into account, one has two relations
\begin{align}
\int\frac{d^{D}l}{i\pi^{D/2}}\frac{(l^0+p^0+\mu_1)}{\mathcal{P}(l,p;m_1,\mu_1)}&=0,\label{EqA1}\\
\int\frac{d^{D}l}{i\pi^{D/2}}\frac{(l^i+p^i)}{\mathcal{P}(l,0;m_1,\mu_1)}&=0,
\end{align}
which lead further to
\begin{align}
\mathscr{A}^{0}(p;m_1,\mu_1;\beta)&=\int\frac{d^{D}l}{i\pi^{D/2}}\frac{-(p^0+\mu_1)}{\mathcal{P}(l,p;m_1,\mu_1)}
\nonumber\\&
=-(\mu_1+p^0)\mathcal{A}_{0}(p;m_1,\mu_1;\beta)=-(\mu_1+p^0)\mathcal{A}_{0}(0;m_1,\mu_1),\\
\mathscr{A}^{i}(p;m_1,\mu_1;\beta)&=\int\frac{d^{D}l}{i\pi^{D/2}}\frac{-p^i}{\mathcal{P}(l,p;m_1,\mu_1)}
\nonumber\\&
=-p^i\mathcal{A}_{0}(p;m_1,\mu_1;\beta)=-p^i\mathcal{A}_{0}(0;m_1,\mu_1;\beta).
\end{align}
In a compact form, the one-point generic one-loop tensor Feynman integral $\mathscr{A}^{\rho}(p;m_1,\mu_1;\beta)$ can be reduced as
\begin{align}
\mathscr{A}^{\rho}(p;m_1,\mu_1;\beta)&\equiv\int\frac{d^{D}l}{i\pi^{D/2}}\frac{l^\rho}{\mathcal{P}(l,p;m_1,\mu_1)}
\nonumber\\&
=-\left(\mu_1\delta^{\rho 0}+p^\rho\right)\mathcal{A}_{0}(0;m_1,\mu_1;\beta)
=-\left(\mu_1u^{\rho}+p^\rho\right)\mathcal{A}_{0}(0;m_1,\mu_1;\beta),\label{EqAFinal}
\end{align}
indicating that the one-point generic one-loop tensor Feynman integral $\mathscr{A}^{\rho}(p;m_1,\mu_1;\beta)$ can be expressed by $\mathcal{A}_{0}(0;m_1,\mu_1;\beta)$.

By contrast, for the relativistic QFTs at zero temperature and zero density, one has
\begin{align}
\mathscr{\tilde{A}}^{\rho}(p;m_1,\mu_{1}=0;\beta=\infty)&=-p^\rho\mathcal{\tilde{A}}_{0}(0;m_1,\mu_{1}=0;\beta=\infty)
\end{align}
in the \emph{conventional} Passarino-Veltman reduction, where the symbols with tilde $\mathcal{\tilde{A}}_{0}(0;m_1,\mu_{1}=0;\beta=\infty)$ and $\mathscr{\tilde{A}}^{\rho}(p;m_1,\mu_{1}=0;\beta=\infty)$ are introduced to denote the quantities at zero temperature. Evidently, $\mathcal{\tilde{A}}_{0}(0;m_1,\mu_{1}=0;\beta=\infty)$ is Lorentz-invariant and $\mathscr{\tilde{A}}^{\rho}(p;m_1,\mu_{1}=0;\beta=\infty)$ is Lorentz covariant.

\subsection{Reduction of two-point generic one-loop tensor Feynman integral $\mathscr{B}^{\rho}(p_1;m_1,\mu_1;m_2,\mu_2;\beta)$}

Before reducing the two-point generic one-loop tensor Feynman integrals $\mathscr{B}^{\rho}(p_1;m_1,\mu_1;m_2,\mu_2;\beta)$ and $\mathscr{B}^{\rho\sigma}(p_1;m_1,\mu_1;m_2,\mu_2;\beta)$, we note that the two-point generic one-loop scalar Feynman integral $\mathcal{B}_{0}(p_1;m_1,\mu_1;m_2,\mu_2;\beta)$ is a function of finite temperature and/or finite density, and depends independently on $p_{1}^{0}$ and $|\boldsymbol{p}_{1}|$, which is not Lorentz invariant. 

Since the two-point generic one-loop tensor Feynman integral $\mathscr{B}^{\rho}(p_1;m_1,\mu_1;m_2,\mu_2;\beta)$ is a rank-one non-Lorentz-covariant tensor, the complete set of tensor structures to expand it can be constructed by a rank-one Lorentz-covariant tensor $p_{1}^{\rho}$ and a rank-one non-Lorentz-covariant tensor $u^{\rho}$. Consequently, $\mathscr{B}^{\rho}(p_1;m_1,\mu_1;m_2,\mu_2;\beta)$ can be reduced as
\begin{align}
&\mathscr{B}^{\rho}(p_1;m_1,\mu_1;m_2,\mu_2;\beta)
\nonumber\\&
=\int\frac{d^{D}l}{i\pi^{D/2}}
\frac{l^\rho}{\mathcal{P}(l,0;m_1,\mu_1)\mathcal{P}(l,p_1;m_2,\mu_2)}
\nonumber\\&
=p_1^\rho\mathcal{B}_1(p_1;m_1,\mu_1;m_2,\mu_2;\beta)
+u^\rho\mathcal{B}_2(p_1;m_1,\mu_1;m_2,\mu_2;\beta)
\end{align}
in the \emph{generalized} Passarino-Veltman reduction. Contracting the two-point generic tensor integral $\mathscr{B}^{\rho}(p_1;m_1,\mu_1;m_2,\mu_2;\beta)$ with $p_{1\rho}$ and $u_{\rho}$ gives rise to two equations,
\begin{align}
G_{B}^{(1)}(p_1)
\left(
  \begin{array}{c}
    \mathcal{B}_1 \\\\
    \mathcal{B}_2 \\
  \end{array}
\right)
&=\left(
  \begin{array}{c}
    \mathcal{F}_1 \\\\
    \mathcal{F}_2 \\
  \end{array}
\right),
\label{BVEqu}
\end{align}
where $G_{B}^{(1)}(p_1)$ is the Gram matrix defined as
\begin{align}
G_{B}^{(1)}(p_1)&=
\left(
  \begin{array}{cc}
    p_1^2 &\hspace{0.3cm} p_1^0 \\\\
    p_1^0 &\hspace{0.3cm} 1 \\
  \end{array}
\right).
\end{align}

In these two equations, $(p_1;m_1,\mu_1;m_2,\mu_2;\beta)$, the arguments of $\mathcal{B}_{1}(p_1;m_1,\mu_1;m_2,\mu_2;\beta)$, $\mathcal{B}_{2}(p_1;m_1,\mu_1;m_2,\mu_2;\beta)$, $\mathcal{F}_{1}(p_1;m_1,\mu_1;m_2,\mu_2;\beta)$, and $\mathcal{F}_{2}(p_1;m_1,\mu_1;m_2,\mu_2;\beta)$ are omitted for short, and $p_{1}^{2}$ is defined as $(p_{1}^{0})^{2}-\boldsymbol{p}_{1}^{2}$. In addition, $\mathcal{F}_1(p_1;m_1,\mu_1;m_2,\mu_2;\beta)$ and $\mathcal{F}_2(p_1;m_1,\mu_1;m_2,\mu_2;\beta)$ are obtained by
\begin{align}
&\mathcal{F}_1(p_1;m_1,\mu_1;m_2,\mu_2;\beta)=p_{1\rho}\mathscr{B}^{\rho}(p_1;m_1,\mu_1;m_2,\mu_2;\beta)
\nonumber\\&
=\int\frac{d^{D}l}{i\pi^{D/2}}
\frac{p_1\cdot l}{\mathcal{P}(l,0;m_1,\mu_1)\mathcal{P}(l,p_1;m_2,\mu_2)},\label{DefF1}\\
&\mathcal{F}_2(p_1;m_1,\mu_1;m_2,\mu_2;\beta)=u_{\rho}\mathscr{B}^{\rho}(p_1;m_1,\mu_1;m_2,\mu_2;\beta)
\nonumber\\&
=\int\frac{d^{D}l}{i\pi^{D/2}}\frac{u\cdot l}{\mathcal{P}(l,0;m_1,\mu_1)\mathcal{P}(l,p_1;m_2,\mu_2)}
\nonumber\\&
=\int\frac{d^{D}l}{i\pi^{D/2}}\frac{l^0}{\mathcal{P}(l,0;m_1,\mu_1)\mathcal{P}(l,p_1;m_2,\mu_2)}.
\end{align}

By solving two equations in (\ref{BVEqu}), the two form factors $\mathcal{B}_{1}(p_1;m_1,\mu_1;m_2,\mu_2;\beta)$ and $\mathcal{B}_{2}(p_1;m_1,\mu_1;m_2,\mu_2;\beta)$ can be expressed in terms of $\mathcal{F}_1(p_1;m_1,\mu_1;m_2,\mu_2;\beta)$ and $\mathcal{F}_2(p_1;m_1,\mu_1;m_2,\mu_2;\beta)$ as
\begin{align}
\left(
  \begin{array}{c}
    \mathcal{B}_1 \\\\
    \mathcal{B}_2 \\
  \end{array}
\right)
&=\frac{1}{\Delta_{B}^{(1)}(p_1)}
\left(
  \begin{array}{cc}
    1 &\hspace{0.3cm} -p_1^0 \\\\
    -p_1^0 &\hspace{0.3cm} p_1^2 \\
  \end{array}
\right)
\left(
  \begin{array}{c}
    \mathcal{F}_1 \\\\
    \mathcal{F}_2 \\
  \end{array}
\right),
\end{align}
where $\Delta_{B}^{(1)}(p_1)=p_1^2-(p_1^0)^2=-\boldsymbol{p}_{1}^{2}$ is the Gram determinant defined by the determinant of the Gram matrix $G_{B}^{(1)}(p_1)$, $\mathcal{F}_{1}(p_1;m_1,\mu_1;m_2,\mu_2;\beta)$ and $\mathcal{F}_2(p_1;m_1,\mu_1;m_2,\mu_2;\beta)$ (see Appendix \ref{App1} for detailed evaluation) are given as
\begin{align}
&\mathcal{F}_1(p_1;m_1,\mu_1;m_2,\mu_2;\beta)
\nonumber\\&
=\frac{\left(\mu_1^2-m_1^2\right)-\left[(\mu_2+p_1^0)^2-m_2^2\right]+\boldsymbol{p}_1^2}{2}
\mathcal{B}_{0}(p_1;m_1,\mu_1;m_2,\mu_2;\beta)
\nonumber\\&
-\left(\mu_2-\mu_1\right)\mathscr{B}^{0}(p_1;m_1,\mu_1;m_2,\mu_2;\beta)
+\frac{\mathcal{A}_{0}(0;m_1,\mu_1;\beta)-\mathcal{A}_{0}(0;m_2,\mu_2;\beta)}{2},
\label{F1Main}
\end{align}
and
\begin{align}
\mathcal{F}_2(p_1;m_1,\mu_1;m_2,\mu_2;\beta)&=\mathscr{B}^{0}(p_1;m_1,\mu_1;m_2,\mu_2;\beta).
\end{align}
Obviously, $\mathcal{F}_{1}(p_1;m_1,\mu_1;m_2,\mu_2;\beta)$ and $\mathcal{F}_{2}(p_1;m_1,\mu_1;m_2,\mu_2;\beta)$ are functions of finite temperature and/or finite density, and they depend independently on $p_{1}^{0}$ and $|\boldsymbol{p}_{1}|$, which are not Lorentz invariant. Consequently, the two-point generic one-loop tensor Feynman integral $\mathscr{B}^{\rho}(p_1;m_1,\mu_1;m_2,\mu_2;\beta)$ can be decomposed as linear combinations
of a Lorentz-covariant tensor $p_1^\rho$ and a non-Lorentz-covariant tensor $u^\rho$, with the non-Lorentz-invariant form factors $\mathcal{B}_1(p_1;m_1,\mu_1;m_2,\mu_2;\beta)$ and $\mathcal{B}_2(p_1;m_1,\mu_1;m_2,\mu_2;\beta)$ being expressed by the generic one-point generic one-loop scalar Feynman integral $\mathcal{A}_{0}(0;m_1,\mu_1;\beta)$, two-point generic one-loop scalar Feynman integral $\mathcal{B}_{0}(p_1;m_1,\mu_1;m_2,\mu_2;\beta)$, and a temporal component of two-point generic one-loop tensor Feynman integral $\mathscr{B}^{0}(p_1;m_1,\mu_1;m_2,\mu_2;\beta)$, which are also non-Lorentz invariant.

By contrast, in the relativistic QFTs at zero temperature and zero density, the $D$-dimensional constant vector $u^{\rho}$ vanishes due to the Lorentz covariance, which leads to 
\begin{align}
\mathscr{\tilde{B}}^{\rho}(p_1;m_1,\mu_1=0;m_2,\mu_2=0;\beta=\infty)
=p_1^\rho\mathcal{\tilde{B}}_1(p_1;m_1,\mu_1=0;m_2,\mu_2=0;\beta=\infty)
\end{align}
in the \emph{conventional} Passarino-Veltman reduction, where the symbols with tilde $\mathcal{\tilde{B}}_{1}(p_1;m_1,\mu_1=0;m_2,\mu_2=0;\beta=\infty)$ and $\mathscr{\tilde{B}}^{\rho}(p_1;m_1,\mu_1=0;m_2,\mu_2=0;\beta=\infty)$ are introduced to denote the quantities at zero temperature. Contracting the two-point genetic one-loop tensor Feynman integral $\mathscr{\tilde{B}}^{\rho}(p_1;m_1,\mu_1=0;m_2,\mu_2=0;\beta=\infty)$ with $p_{1\rho}$ gives rise to an equation, whose solution is evidently Lorentz invariant, namely,
\begin{align}
\mathcal{\tilde{B}}_1(p_1;m_1,\mu_1=0;m_2,\mu_2=0;\beta=\infty)
&=\frac{1}{p_1^{2}}\mathcal{\tilde{F}}_1(p_1;m_1,\mu_1=0;m_2,\mu_2=0;\beta=\infty),
\end{align}
where
\begin{align}
&\mathcal{\tilde{F}}_1(p_1;m_1,\mu_1=0;m_2,\mu_2=0;\beta=\infty)
=p_{1\rho}\mathscr{\tilde{B}}^{\rho}(p_1;m_1,\mu_1=0;m_2,\mu_2=0;\beta=\infty)
\nonumber\\&
=\frac{m_2^2-m_1^{2}-p_1^2}{2}\mathcal{\tilde{B}}_{0}(p_1;m_1,\mu_1=0;m_2,\mu_2=0;\beta=\infty)
\nonumber\\&\hspace{0.4cm}
+\frac{\mathcal{\tilde{A}}_{0}(0;m_1,\mu_1=0;\beta=\infty)-\mathcal{\tilde{A}}_{0}(0;m_2,\mu_2=0;\beta=\infty)}{2},
\end{align}
which agrees exactly with the results presented in the Appendix A of Ref. \cite{Ellis}. It is noted that $\mathcal{\tilde{B}}_{0}(p_1;m_1,\mu_1=0;m_2,\mu_2;\beta=\infty)$ and $\mathcal{\tilde{A}}_{0}(0;m,\mu=0;\beta=\infty)$ are Lorentz invariant. Hence, $\mathcal{\tilde{B}}_1(p_1;m_1,\mu_1=0;m_2,\mu_2=0;\beta=\infty)$ is Lorentz invariant and $\mathscr{\tilde{B}}^{\rho}(p_1;m_1,\mu_1=0;m_2,\mu_2=0;\beta=\infty)$ is Lorentz covariant. It is stressed that $\mathcal{\tilde{F}}_{1}(p_1;m_1,\mu_1=0;m_2,\mu_2=0;\beta=\infty)$ can be obtained by directly setting $\mu_{1}=\mu_{2}=0$ and $T=0$ in Eq.(\ref{F1Main}).

\subsection{Reduction of two-point generic one-loop tensor Feynman integral $\mathscr{B}^{\rho\sigma}(p_1;m_1,\mu_1;m_2,\mu_2;\beta)$}

Because the two-point generic one-loop tensor Feynman integral $\mathscr{B}^{\rho\sigma}(p_1;m_1,\mu_1;m_2,\mu_2;\beta)$ is a symmetric rank-two tensor, the complete set of tensor structures to expand it can be constructed by the symmetric rank-two Lorentz-covariant tensors $g^{\rho\sigma}$ and $p_{1}^{\rho}p_{1}^{\sigma}$, and the symmetric rank-two non-Lorentz-covariant tensors $u^{\rho}u^{\sigma}$ and $\left(p_{1}^{\rho}u^{\sigma}+p_{1}^{\sigma}u^{\rho}\right)$. Consequently, $\mathscr{B}^{\rho\sigma}(p_1;m_1,\mu_1;m_2,\mu_2;\beta)$ can be reduced as
\begin{align}
&\mathscr{B}^{\rho\sigma}(p_1;m_1,\mu_1;m_2,\mu_2;\beta)
\nonumber\\&
=\int\frac{d^{D}l}{i\pi^{D/2}}
\frac{l^{\rho}l^{\sigma}}{\mathcal{P}(l,0;m_1,\mu_1)\mathcal{P}(l,p_1;m_2,\mu_2)}
\nonumber\\&
=g^{\rho\sigma}\mathcal{B}_{00}(p_1;m_1,\mu_1;m_2,\mu_2;\beta)
+p_{1}^{\rho}p_{1}^{\sigma}\mathcal{B}_{11}(p_1;m_1,\mu_1;m_2,\mu_2;\beta)
\nonumber\\&\hspace{0.4cm}
+\left(p_{1}^{\rho}u^{\sigma}+p_{1}^{\sigma}u^{\rho}\right)\mathcal{B}_{12}(p_1;m_1,\mu_1;m_2,\mu_2;\beta)
+u^{\rho}u^{\sigma}\mathcal{B}_{22}(p_1;m_1,\mu_1;m_2,\mu_2;\beta)
\end{align}
in the \emph{generalized} Passarino-Veltman reduction, where $\mathcal{B}_{ab}(p_1;m_1,\mu_1;m_2,\mu_2;\beta)=\mathcal{B}_{ba}(p_1;m_1,\mu_1;m_2,\mu_2;\beta)$. 

Contracting the two-point genetic one-loop tensor Feynman integral $\mathscr{B}^{\rho\sigma}(p_1;m_1,\mu_1;m_2,\mu_2;\beta)$ with $g_{\rho\sigma}$, $p_{1\rho}p_{1\sigma}$, $p_{1\rho}u_{\sigma}$, and $u_{\rho}u_{\sigma}$ gives rise to four equations,
\begin{align}
G_{B}^{(2)}(p_1)
\left(
  \begin{array}{c}
    \mathcal{B}_{00} \\\\
    \mathcal{B}_{11} \\\\
    \mathcal{B}_{12} \\\\
    \mathcal{B}_{22} \\
  \end{array}
\right)
=\left(
  \begin{array}{c}
    \mathcal{F}_{00} \\\\
    \mathcal{F}_{11} \\\\
    \mathcal{F}_{12} \\\\
    \mathcal{F}_{22} \\
  \end{array}
\right),
\label{BTEqu}
\end{align}
where $G_{B}^{(2)}(p_1)$ is the Gram matrix defined as
\begin{align}
G_{B}^{(2)}(p_1)&=
\left(
  \begin{array}{cccc}
    D &\hspace{0.3cm} p_1^2 &\hspace{0.3cm} 2p_1^0 &\hspace{0.3cm} 1 \\\\
    p_1^2 &\hspace{0.3cm} (p_1^2)^2 &\hspace{0.3cm} 2p_1^0 p_1^2 &\hspace{0.3cm} (p_1^0)^2 \\\\
    p_1^0 &\hspace{0.3cm} p_1^2 p_1^0 &\hspace{0.3cm} (p_1^0)^2+p_1^2 &\hspace{0.3cm} p_1^0 \\\\
    1 &\hspace{0.3cm} (p_1^0)^2 &\hspace{0.3cm} 2p_1^0 &\hspace{0.3cm} 1 \\
  \end{array}
\right).
\end{align}
The arguments $(p_1;m_1,\mu_1;m_2,\mu_2;\beta)$ in $\mathcal{B}_{00}(p_1;m_1,\mu_1;m_2,\mu_2;\beta)$,  $\mathcal{B}_{11}(p_1;m_1,\mu_1;m_2,\mu_2;\beta)$, $\mathcal{B}_{12}(p_1;m_1,\mu_1;m_2,\mu_2;\beta)$, $\mathcal{B}_{22}(p_1;m_1,\mu_1;m_2,\mu_2;\beta)$, $\mathcal{F}_{00}(p_1;m_1,\mu_1;m_2,\mu_2;\beta)$, $\mathcal{F}_{11}(p_1;m_1,\mu_1;m_2,\mu_2;\beta)$, $\mathcal{F}_{12}(p_1;m_1,\mu_1;m_2,\mu_2;\beta)$, and $\mathcal{F}_{22}(p_1;m_1,\mu_1;m_2,\mu_2;\beta)$ are omitted for short. In addition, $\mathcal{F}_{00}(p_1;m_1,\mu_1;m_2,\mu_2;\beta)$, $\mathcal{F}_{11}(p_1;m_1,\mu_1;m_2,\mu_2;\beta)$, $\mathcal{F}_{12}(p_1;m_1,\mu_1;m_2,\mu_2;\beta)$, and $\mathcal{F}_{22}(p_1;m_1,\mu_1;m_2,\mu_2;\beta)$ are obtained by
\begin{align}
&\mathcal{F}_{00}(p_1;m_1,\mu_1;m_2,\mu_2;\beta)=g_{\rho\sigma}\mathscr{B}^{\rho\sigma}(p_1;m_1,\mu_1;m_2,\mu_2;\beta)
\nonumber\\&
=\int\frac{d^{D}l}{i\pi^{D/2}}\frac{l^{2}}{\mathcal{P}(l,0;m_1,\mu_1)\mathcal{P}(l,p_1;m_2,\mu_2)},\label{DefF00}\\
&\mathcal{F}_{11}(p_1;m_1,\mu_1;m_2,\mu_2;\beta)=p_{1\rho}p_{1\sigma}\mathscr{B}^{\rho\sigma}(p_1;m_1,\mu_1;m_2,\mu_2;\beta)
\nonumber\\&
=\int\frac{d^{D}l}{i\pi^{D/2}}\frac{(p_{1}\cdot l)^2}{\mathcal{P}(l,0;m_1,\mu_1)\mathcal{P}(l,p_1;m_2,\mu_2)},\label{DefF11}\\
&\mathcal{F}_{12}(p_1;m_1,\mu_1;m_2,\mu_2;\beta)=p_{1\rho}u_{\sigma}\mathscr{B}^{\rho\sigma}(p_1;m_1,\mu_1;m_2,\mu_2;\beta)
\nonumber\\&
=\int\frac{d^{D}l}{i\pi^{D/2}}\frac{(u\cdot l)(p_{1}\cdot l)}{\mathcal{P}(l,0;m_1,\mu_1)\mathcal{P}(l,p_1;m_2,\mu_2)}
\nonumber\\&
=\int\frac{d^{D}l}{i\pi^{D/2}}\frac{l^{0}(p_{1}\cdot l)}{\mathcal{P}(l,0;m_1,\mu_1)\mathcal{P}(l,p_1;m_2,\mu_2)},\label{DefF12}\\
&\mathcal{F}_{22}(p_1;m_1,\mu_1;m_2,\mu_2;\beta)=u_{\rho}u_{\sigma}\mathscr{B}^{\rho\sigma}(p_1;m_1,\mu_1;m_2,\mu_2;\beta)
\nonumber\\&
=\int\frac{d^{D}l}{i\pi^{D/2}}\frac{(u\cdot l)(u\cdot l)}{\mathcal{P}(l,0;m_1,\mu_1)\mathcal{P}(l,p_1;m_2,\mu_2)}
\nonumber\\&
=\int\frac{d^{D}l}{i\pi^{D/2}}\frac{l^{0}l^{0}}{\mathcal{P}(l,0;m_1,\mu_1)\mathcal{P}(l,p_1;m_2,\mu_2)}\label{DefF22}.
\end{align}

By solving four equations in (\ref{BTEqu}), we can express the four form factors $\mathcal{B}_{00}(p_1;m_1,\mu_1;m_2,\mu_2;\beta)$, $\mathcal{B}_{11}(p_1;m_1,\mu_1;m_2,\mu_2;\beta)$, $\mathcal{B}_{12}(p_1;m_1,\mu_1;m_2,\mu_2;\beta)$, and $\mathcal{B}_{22}(p_1;m_1,\mu_1;m_2,\mu_2;\beta)$ in terms of $\mathcal{F}_{00}(p_1;m_1,\mu_1;m_2,\mu_2;\beta)$, $\mathcal{F}_{11}(p_1;m_1,\mu_1;m_2,\mu_2;\beta)$, $\mathcal{F}_{12}(p_1;m_1,\mu_1;m_2,\mu_2;\beta)$, and $\mathcal{F}_{22}(p_1;m_1,\mu_1;m_2,\mu_2;\beta)$ as
\begin{align}
\left(
  \begin{array}{c}
    \mathcal{B}_{00} \\\\\\
    \mathcal{B}_{11} \\\\\\
    \mathcal{B}_{12} \\\\\\
    \mathcal{B}_{22} \\
  \end{array}
\right)
=
\frac{\left[p_1^2-(p_1^0)^2\right]^2}{\Delta_{B}^{(2)}(p_1)}
\left(
  \begin{array}{cccc}
  p_1^2-(p_1^0)^2 &\hspace{0.3cm} -1 &\hspace{0.3cm} 2p_1^0 &\hspace{0.3cm} -p_1^2 \\\\
    -1 &\hspace{0.3cm} \frac{(D-1)}{p_1^2-(p_1^0)^2} &\hspace{0.3cm} -\frac{2(D-1)p_1^0}{p_1^2-(p_1^0)^2} &\hspace{0.3cm} \frac{(D-2)(p_1^0)^2+p_1^2}{p_1^2-(p_1^0)^2} \\\\\\
  p_1^0 &\hspace{0.3cm} -\frac{(D-1)p_1^0}{p_1^2-(p_1^0)^2} &\hspace{0.3cm} 
  \frac{(D-2)p_1^2+D(p_1^0)^2}{p_1^2-(p_1^0)^2} &\hspace{0.3cm} -\frac{(D-1)p_1^0 p_1^2}{p_1^2-(p_1^0)^2} \\\\
   -p_1^2  &\hspace{0.3cm} \frac{(D-2)(p_1^0)^2+p_1^2}{p_1^2-(p_1^0)^2} &\hspace{0.3cm}
  -\frac{2(D-1)p_1^0 p_1^2}{p_1^2-(p_1^0)^2} &\hspace{0.3cm} \frac{(D-1)(p_1^2)^2}{p_1^2-(p_1^0)^2} \\
  \end{array}
\right)
\left(
  \begin{array}{c}
    \mathcal{F}_{00} \\\\\\
    \mathcal{F}_{11} \\\\\\
    \mathcal{F}_{12} \\\\\\
    \mathcal{F}_{22} \\
  \end{array}
\right),
\end{align} 
where $\Delta_{B}^{(2)}(p_1)=(D-2)\left[p_1^2-(p_1^0)^2\right]^3$ with $p_1^2-(p_1^0)^2=-\boldsymbol{p}_{1}^{2}$ is the Gram determinant defined by the determinant of the Gram matrix $G_{b}^{(2)}(p_1)$. These solutions indicate that the \emph{generalized} Passarino-Veltman reduction is singular when $D=2$, which originates from the combination of the Lorentz-covariance breaking and the $(1+1)$-dimensional spacetime. Physically, it is unnecessary to further reduce the one-loop tensor Feynman integrals in the $(1+1)$-dimensional spacetime, because the space component and the temporal component are not equivalent to each other when the Lorentz covariance is explicitly broken. 

Four axillary functions $\mathcal{F}_{00}(p_1;m_1,\mu_1;m_2,\mu_2;\beta)$, $\mathcal{F}_{11}(p_1;m_1,\mu_1;m_2,\mu_2;\beta)$, $\mathcal{F}_{12}(p_1;m_1,\mu_1;m_2,\mu_2;\beta)$, and $\mathcal{F}_{22}(p_1;m_1,\mu_1;m_2,\mu_2;\beta)$ (see Appendix \ref{App2} for detailed evaluation) can be expressed as
\begin{align}
&\mathcal{F}_{00}(p_1;m_1,\mu_1;m_2,\mu_2;\beta)
\nonumber\\&
=-\left(\mu_1^2-m_1^2\right)\mathcal{B}_{0}(p_1;m_1,\mu_1;m_2,\mu_2;\beta)
-2\mu_1\mathscr{B}^{0}(p_1;m_1,\mu_1;m_2,\mu_2;\beta)+\mathcal{A}_{0}(0;m_2,\mu_2;\beta),
\label{F00Main}
\end{align}
\begin{align}
&\mathcal{F}_{11}(p_1;m_1,\mu_1;m_2,\mu_2;\beta)
\nonumber\\&
=\frac{\left\{\left(\mu_1^2-m_1^2\right)-\left[(\mu_2+p_1^0)^2-m_2^2\right]+\boldsymbol{p}_1^2\right\}^2}{4}
\mathcal{B}_{0}(p_1;m_1,\mu_1;m_2,\mu_2;\beta)
\nonumber\\&\hspace{0.4cm}
+\left(\mu_2-\mu_1\right)^2\mathscr{B}^{00}(p_1;m_1,\mu_1;m_2,\mu_2;\beta)
\nonumber\\&\hspace{0.4cm}
-\left(\mu_2-\mu_1\right)\left\{\left(\mu_1^2-m_1^2\right)
-\left[(\mu_2+p_1^0)^2-m_2^2\right]+\boldsymbol{p}_1^2\right\}
\mathscr{B}^{0}(p_1;m_1,\mu_1;m_2,\mu_2;\beta)
\nonumber\\&\hspace{0.4cm}
+\frac{\left(m_2^2-m_1^2-p_1^2\right)-\left(\mu_1-\mu_2\right)^2-2p_1^0\left(\mu_1+\mu_2\right)}{4}
\mathcal{A}_{0}(0;m_1,\mu_1;\beta)
\nonumber\\&\hspace{0.4cm}
+\frac{\left(m_1^2-m_2^2+3p_1^2\right)-\left(\mu_1-\mu_2\right)^2+2p_1^0\left(\mu_1+\mu_2\right)}{4}
\mathcal{A}_{0}(0;m_2,\mu_2;\beta),
\label{F11Main}
\end{align}
\begin{align}
&\mathcal{F}_{12}(p_1;m_1,\mu_1;m_2,\mu_2;\beta)
\nonumber\\&
=\frac{\left(\mu_1^2-m_1^2\right)
-\left[(\mu_2+p_1^0)^2-m_2^2\right]+\boldsymbol{p}_1^2}{2}\mathscr{B}^{0}(p_1;m_1,\mu_1;m_2,\mu_2;\beta)
\nonumber\\&\hspace{0.4cm}
-\left(\mu_2-\mu_1\right)\mathscr{B}^{00}(p_1;m_1,\mu_1;m_2,\mu_2;\beta)
\nonumber\\&\hspace{0.4cm}
+\frac{-\mu_1\mathcal{A}_0(0;m_1,\mu_1;\beta)+(\mu_2+p_{1}^{0})\mathcal{A}_0(0;m_2,\mu_2;\beta)}{2},
\end{align}
and
\begin{align}
\mathcal{F}_{22}(p_1;m_1,\mu_1;m_2,\mu_2;\beta)&=\mathscr{B}^{00}(p_1;m_1,\mu_1;m_2,\mu_2;\beta).
\end{align}
Obviously, $\mathcal{F}_{00}(p_1;m_1,\mu_1;m_2,\mu_2;\beta)$, $\mathcal{F}_{11}(p_1;m_1,\mu_1;m_2,\mu_2;\beta)$, $\mathcal{F}_{12}(p_1;m_1,\mu_1;m_2,\mu_2;\beta)$, and $\mathcal{F}_{22}(p_1;m_1,\mu_1;m_2,\mu_2;\beta)$ are functions of finite temperature and/or finite density, and they depend independently on $p_{1}^{0}$ and $|\boldsymbol{p}_{1}|$, which are not Lorentz invariant. Consequently, the two-point generic one-loop tensor Feynman integral $\mathscr{B}^{\rho\sigma}(p_1;m_1,\mu_1;m_2,\mu_2;\beta)$ can be decomposed as linear combinations of two Lorentz-covariant tensors $g^{\rho\sigma}$ and $p_{1}^{\rho}p_{1}^{\sigma}$, and two Lorentz-covariant tensors $u^{\rho}u^{\sigma}$ and $\left(p_{1}^{\rho}u^{\sigma}+p_{1}^{\sigma}u^{\rho}\right)$, with the non-Lorentz-invariant form factors $\mathcal{B}_{00}(p_1;m_1,\mu_1;m_2,\mu_2;\beta)$, $\mathcal{B}_{11}(p_1;m_1,\mu_1;m_2,\mu_2;\beta)$, $\mathcal{B}_{12}(p_1;m_1,\mu_1;m_2,\mu_2;\beta)$, and $\mathcal{B}_{22}(p_1;m_1,\mu_1;m_2,\mu_2;\beta)$ being expressed by the one-point generic one-loop scalar Feynman integral $\mathcal{A}_{0}(0;m_1,\mu_1;\beta)$, two-point generic one-loop scalar Feynman integral $\mathcal{B}_{0}(p_1;m_1,\mu_1;m_2,\mu_2;\beta)$, and two temporal components of two-point generic one-loop tensor Feynman integrals $\mathscr{B}^{0}(p_1;m_1,\mu_1;m_2,\mu_2;\beta)$ and $\mathscr{B}^{00}(p_1;m_1,\mu_1;m_2,\mu_2;\beta)$, which are also non-Lorentz invariant.

By contrast, for the relativistic QFTs at zero temperature and zero density, the $D$-dimensional constant vectors $u^{\rho}$ and $u^{\sigma}$ vanish due to the Lorentz covariance, which leads to 
\begin{align}
&\mathscr{\tilde{B}}^{\rho\sigma}(p_1;m_1,\mu_1=0;m_2,\mu_2=0;\beta=\infty)
\nonumber\\&
=g^{\rho\sigma}\mathcal{\tilde{B}}_{00}(p_1;m_1,\mu_1=0;m_2,\mu_2=0;\beta=\infty)
+p_1^{\rho}p_1^{\sigma}\mathcal{\tilde{B}}_{11}(p_1;m_1,\mu_1=0;m_2,\mu_2=0;\beta=\infty)
\end{align}
in the \emph{conventional} Passarino-Veltman reduction, where the symbols with tilde $\mathcal{\tilde{B}}_{00}(p_1;m_1,\mu_1=0;m_2,\mu_2=0;\beta=\infty)$, $\mathcal{\tilde{B}}_{11}(p_1;m_1,\mu_1=0;m_2,\mu_2=0;\beta=\infty)$, and $\mathscr{\tilde{B}}^{\rho\sigma}(p_1;m_1,\mu_1=0;m_2,\mu_2=0;\beta=\infty)$ are introduced to denote the quantities at zero temperature. Contracting the two-point genetic one-loop tensor Feynman integral $\mathscr{\tilde{B}}^{\rho\sigma}(p_1;m_1,\mu_1=0;m_2,\mu_2=0;\beta=\infty)$ with $g_{\rho\sigma}$ and $p_{1\rho}p_{1\sigma}$ gives rise to two equations, 
\begin{align}
\tilde{G}_{B}^{(2)}(p_1)
\left(
  \begin{array}{c}
    \mathcal{\tilde{B}}_{00} \\\\\\
    \mathcal{\tilde{B}}_{11} \\
  \end{array}
\right)
&=\left(
  \begin{array}{c}
    \mathcal{\tilde{F}}_{00} \\\\\\
    \mathcal{\tilde{F}}_{11} \\
  \end{array}
\right),
\end{align}
where $\tilde{G}_{B}^{(2)}(p_1)$ is the Gram matrix defined as 
\begin{align}
\tilde{G}_{B}^{(2)}(p_1)&=
\left(
  \begin{array}{cc}
  D &\hspace{0.3cm} p_1^2\\\\\\
  p_1^2  &\hspace{0.3cm} (p_1^2)^{2}  \\
  \end{array}
\right).
\end{align}
The arguments $(p_1;m_1,\mu_1=0;m_2,\mu_2=0;\beta)$ in $\mathcal{\tilde{B}}_{00}(p_1;m_1,\mu_1=0;m_2,\mu_2=0;\beta=\infty)$, $\mathcal{\tilde{B}}_{11}(p_1;m_1,\mu_1=0;m_2,\mu_2=0;\beta=\infty)$, $\mathcal{\tilde{F}}_{00}(p_1;m_1,\mu_1=0;m_2,\mu_2=0;\beta=\infty)$, and $\mathcal{\tilde{F}}_{11}(p_1;m_1,\mu_1=0;m_2,\mu_2=0;\beta=\infty)$ are omitted for short. The solutions to these two equations are given as
\begin{align}
\left(
  \begin{array}{c}
    \mathcal{\tilde{B}}_{00} \\\\\\
    \mathcal{\tilde{B}}_{11} \\
  \end{array}
\right)
=
\frac{1}{\tilde{\Delta}_{B}^{(2)}(p_1)}
\left(
  \begin{array}{cc}
  (p_1^2)^{2} &\hspace{0.3cm} -p_1^2\\\\\\
  -p_1^2  & D\hspace{0.3cm}  \\
  \end{array}
\right)
\left(
  \begin{array}{c}
    \mathcal{\tilde{F}}_{00} \\\\\\
    \mathcal{\tilde{F}}_{11} \\
  \end{array}
\right).
\end{align}
where $\tilde{\Delta}_{B}^{(2)}(p_1)=(D-1)(p_1^2)^{2}$ is the Gram determinant defined by the determinant of the Gram matrix $\tilde{G}_{B}^{(2)}(p_1)$. These Lorentz-invariant solutions also indicate that the \emph{conventional} Passarino-Veltman reduction is singular when $D=1$, which originates from the $(1+0)$-dimensional spacetime. Physically, it is unnecessary to further reduce the one-loop tensor Feynman integrals in the $(1+0)$-dimensional spacetime, because there is no spatial component when $D=1$.

For the relativistic QFTs at zero temperature and zero density, it is obvious that
\begin{align}
&\mathcal{\tilde{F}}_{00}(p_1;m_1,\mu_1=0;m_2,\mu_2=0;\beta=\infty)
\nonumber\\&
=m_1^2\mathcal{\tilde{B}}_{0}(p_1;m_1,\mu_1=0;m_2,\mu_2=0;\beta)+\mathcal{\tilde{A}}_{0}(0;m_2,\mu_2;\beta=\infty),
\end{align}
and
\begin{align}
&\mathcal{\tilde{F}}_{11}(p_1;m_1,\mu_1=0;m_2,\mu_2=0;\beta=\infty)
\nonumber\\&
=\frac{\left(m_2^2-m_1^2-p_1^2\right)^2}{4}\mathcal{\tilde{B}}_{0}(p_1;m_1,\mu_1=0;m_2,\mu_2=0;\beta=\infty)
\nonumber\\&\hspace{0.4cm}
+\frac{m_2^2-m_1^2-p_1^2}{4}\mathcal{\tilde{A}}_{0}(0;m_1,\mu_1=0;\beta=\infty)
+\frac{m_1^2-m_2^2+3p_1^2}{4}\mathcal{\tilde{A}}_{0}(0;m_2,\mu_2=0;\beta=\infty),
\end{align}
which agree exactly with the results presented in the Appendix A of Ref. \cite{Ellis}. It is stressed that $\mathcal{\tilde{F}}_{00}(p_1;m_1,\mu_1=0;m_2,\mu_2=0;\beta=\infty)$ and $\mathcal{\tilde{F}}_{11}(p_1;m_1,\mu_1=0;m_2,\mu_2=0;\beta=\infty)$ can be obtained by directly setting $\mu_{1}=\mu_{2}=0$ and $T=0$ in Eqs.(\ref{F00Main}) and (\ref{F11Main}), respectively.

\subsection{Reduction for three-point generic one-loop tensor Feynman integral $\mathscr{C}^{\rho}(p_1,p_2;m_1,\mu_1;m_2,\mu_2;m_3,\mu_3;\beta)$}

Before reducing the three-point generic one-loop tensor Feynman integrals $\mathscr{C}^{\rho}(p_1,p_2;m_1,\mu_1;m_2,\mu_2;m_3,\mu_3;\beta)$, $\mathscr{C}^{\rho\sigma}(p_1,p_2;m_1,\mu_1;m_2,\mu_2;m_3,\mu_3;\beta)$, and $\mathscr{C}^{\rho\sigma\tau}(p_1,p_2;m_1,\mu_1;m_2,\mu_2;m_3,\mu_3;\beta)$, we stress that the three-point generic one-loop scalar Feynman integral $\mathcal{C}_{0}(p_1,p_2;m_1,\mu_1;m_2,\mu_2;m_3,\mu_3;\beta)$ is a non-Lorentz-invariant function depending independently on $p_{1}^{0}$, $|\boldsymbol{p}_{1}|$, $p_{2}^{0}$, and $|\boldsymbol{p}_{2}|$ at finite temperature and/or finite density. 

Since the three-point generic one-loop tensor Feynman integral $\mathscr{C}^{\rho}(p_1,p_2;m_1,\mu_1;m_2,\mu_2;m_3,\mu_3;\beta)$ is a rank-one non-Lorentz-covariant tensor, the complete set of tensor structures to expand it can be constructed by two rank-one Lorentz-covariant tensors $p_{1}^{\rho}$ and $p_{2}^{\rho}$, and a rank-one non-Lorentz-covariant tensor $u^{\rho}$. Consequently, $\mathscr{C}^{\rho}(p_1,p_2;m_1,\mu_1;m_2,\mu_2;m_3,\mu_3;\beta)$ can be reduced as
\begin{align}
&\mathscr{C}^{\rho}(p_1,p_2;m_1,\mu_1;m_2,\mu_2;m_3,\mu_3;\beta)
\nonumber\\&
=\int\frac{d^{D}l}{i\pi^{D/2}}
\frac{l^{\rho}}{\mathcal{P}(l,0;m_1,\mu_1)\mathcal{P}(l,p_1;m_2,\mu_2)\mathcal{P}(l,p_1+p_2;m_3,\mu_3)}
\nonumber\\&
=p_{1}^{\rho}\mathcal{C}_{1}(p_1,p_2;m_1,\mu_1;m_2,\mu_2;m_3,\mu_3;\beta)
+p_{2}^{\rho}\mathcal{C}_{2}(p_1,p_2;m_1,\mu_1;m_2,\mu_2;m_3,\mu_3;\beta)
\nonumber\\&\hspace{0.4cm}
+u^{\rho}\mathcal{C}_{3}(p_1,p_2;m_1,\mu_1;m_2,\mu_2;m_3,\mu_3;\beta).
\end{align}

Contracting the two-point genetic tensor integral $\mathscr{C}^{\rho}(p_1,p_2;m_1,\mu_1;m_2,\mu_2;m_3,\mu_3;\beta)$ with $p_{1\rho}$, $p_{2\rho}$, and $u_{\rho}$ gives rise to three equations,
\begin{align}
G_{C}^{(1)}(p_1,p_2)
\left(
  \begin{array}{c}
    \mathcal{C}_{1} \\\\
    \mathcal{C}_{2} \\\\
    \mathcal{C}_{3} \\
  \end{array}
\right)
=\left(
  \begin{array}{c}
    \mathcal{K}_{1} \\\\
    \mathcal{K}_{2} \\\\
    \mathcal{K}_{3} \\
  \end{array}
\right),
\label{COLTFI}
\end{align}
where $G_{C}^{(1)}(p_1,p_2)$ is the Gram matrix defined by
\begin{align}
G_{C}^{(1)}(p_1,p_2)
&=\left(
  \begin{array}{ccc}
    p_1^2 & p_1\cdot p_2 & p_1^0\\\\
    p_1\cdot p_2 & p_2^2 & p_2^0 \\\\
    p_1^0 & p_2^0 & 1 \\
  \end{array}
\right).
\end{align}

In these three equations, the arguments $(p_1,p_2;m_1,\mu_1;m_2,\mu_2;m_3,\mu_3;\beta)$ in $\mathcal{C}_{1}(p_1,p_2;m_1,\mu_1;m_2,\mu_2;m_3,\mu_3;\beta)$, $\mathcal{C}_{2}(p_1,p_2;m_1,\mu_1;m_2,\mu_2;m_3,\mu_3;\beta)$, $\mathcal{C}_{3}(p_1,p_2;m_1,\mu_1;m_2,\mu_2;m_3,\mu_3;\beta)$, $\mathcal{K}_{1}(p_1,p_2;m_1,\mu_1;m_2,\mu_2;m_3,\mu_3;\beta)$, $\mathcal{K}_{2}(p_1,p_2;m_1,\mu_1;m_2,\mu_2;m_3,\mu_3;\beta)$, and $\mathcal{K}_{3}(p_1,p_2;m_1,\mu_1;m_2,\mu_2;m_3,\mu_3;\beta)$ are omitted for short. In addition, $\mathcal{K}_{1}(p_1,p_2;m_1,\mu_1;m_2,\mu_2;m_3,\mu_3;\beta)$, $\mathcal{K}_{2}(p_1,p_2;m_1,\mu_1;m_2,\mu_2;m_3,\mu_3;\beta)$, and $\mathcal{K}_{3}(p_1,p_2;m_1,\mu_1;m_2,\mu_2;m_3,\mu_3;\beta)$ are defined by
\begin{align}
&\mathcal{K}_1(p_1,p_2;m_1,\mu_1;m_2,\mu_2;m_3,\mu_3;\beta)=p_{1\rho}\mathscr{C}^{\rho}(p_1,p_2;m_1,\mu_1;m_2,\mu_2;m_3,\mu_3;\beta)
\nonumber\\&
=\int\frac{d^{D}l}{i\pi^{D/2}}
\frac{p_{1}\cdot l}{\mathcal{P}(l,0;m_1,\mu_1)\mathcal{P}(l,p_1;m_2,\mu_2)\mathcal{P}(l,p_1+p_2;m_3,\mu_3)},
\label{DefG1}\\
&\mathcal{K}_2(p_1,p_2;m_1,\mu_1;m_2,\mu_2;m_3,\mu_3;\beta)=p_{2\rho}\mathscr{C}^{\rho}(p_1,p_2;m_1,\mu_1;m_2,\mu_2;m_3,\mu_3;\beta)
\nonumber\\&
=\int\frac{d^{D}l}{i\pi^{D/2}}\frac{p_{2}\cdot l}{\mathcal{P}(l,0;m_1,\mu_1)\mathcal{P}(l,p_1;m_2,\mu_2)\mathcal{P}(l,p_1+p_2;m_3,\mu_3)},
\label{DefG2}\\
&\mathcal{K}_3(p_1,p_2;m_1,\mu_1;m_2,\mu_2;m_3,\mu_3;\beta)=u_{\rho}\mathscr{C}^{\rho}(p_1,p_2;m_1,\mu_1;m_2,\mu_2;m_3,\mu_3;\beta)
\nonumber\\&
=\int\frac{d^{D}l}{i\pi^{D/2}}
\frac{u\cdot l}{\mathcal{P}(l,0;m_1,\mu_1)\mathcal{P}(l,p_1;m_2,\mu_2)\mathcal{P}(l,p_1+p_2;m_3,\mu_3)}
\nonumber\\&
=\int\frac{d^{D}l}{i\pi^{D/2}}
\frac{l^{0}}{\mathcal{P}(l,0;m_1,\mu_1)\mathcal{P}(l,p_1;m_2,\mu_2)\mathcal{P}(l,p_1+p_2;m_3,\mu_3)}.
\label{DefG3}
\end{align}

By solving three equations in (\ref{COLTFI}), the form factors $\mathcal{C}_{1}(p_1,p_2;m_1,\mu_1;m_2,\mu_2;m_3,\mu_3;\beta)$, $\mathcal{C}_{2}(p_1,p_2;m_1,\mu_1;m_2,\mu_2;m_3,\mu_3;\beta)$, and $\mathcal{C}_{3}(p_1,p_2;m_1,\mu_1;m_2,\mu_2;m_3,\mu_3;\beta)$ can be expressed in terms of $\mathcal{K}_{1}(p_1,p_2;m_1,\mu_1;m_2,\mu_2;m_3,\mu_3;\beta)$, $\mathcal{K}_{2}(p_1,p_2;m_1,\mu_1;m_2,\mu_2;m_3,\mu_3;\beta)$, and $\mathcal{K}_{3}(p_1,p_2;m_1,\mu_1;m_2,\mu_2;m_3,\mu_3;\beta)$ as
\begin{align}
\left(
  \begin{array}{c}
    \mathcal{C}_{1} \\\\
    \mathcal{C}_{2} \\\\
    \mathcal{C}_{3} \\
  \end{array}
\right)
&=\frac{1}{\Delta_{C}^{(1)}(p_{1},p_{2})}
\left(
  \begin{array}{ccc}
    p_2^2-(p_2^0)^2 &\hspace{0.3cm} p_1^0p_2^0-(p_1\cdot p_2) &\hspace{0.3cm} (p_1\cdot p_2) p_2^0-p_1^0p_2^2 \\\\
    p_1^0p_2^0-(p_1\cdot p_2) &\hspace{0.3cm} p_1^2-(p_1^0)^2 &\hspace{0.3cm} (p_1\cdot p_2) p_1^0-p_1^2p_2^0 \\\\
    (p_1\cdot p_2) p_2^0-p_1^0p_2^2 &\hspace{0.3cm} (p_1\cdot p_2) p_1^0-p_1^2p_2^0  &\hspace{0.3cm} p_{1}^{2}p_{2}^{2}-(p_1\cdot p_2)^{2} \\
  \end{array}
\right)
\left(
  \begin{array}{c}
    \mathcal{G}_{1} \\\\
    \mathcal{G}_{2} \\\\
    \mathcal{G}_{3} \\
  \end{array}
\right),
\end{align}
where $\Delta_{C}^{(1)}(p_{1},p_{2})$ is defined by $\Delta_{c}^{(1)}(p_{1},p_{2})=p_{1}^{2}p_{2}^{2}-(p_1\cdot p_2)^{2}
-\left(p_{1} p_{2}^{0}-p_{1}^{0} p_{2}\right)^{2}$, 
$\mathcal{K}_{1}(p_1,p_2;m_1,\mu_1;m_2,\mu_2;m_3,\mu_3;\beta)$, $\mathcal{K}_{2}(p_1,p_2;m_1,\mu_1;m_2,\mu_2;m_3,\mu_3;\beta)$, and $\mathcal{K}_{3}(p_1,p_2;m_1,\mu_1;m_2,\mu_2;m_3,\mu_3;\beta)$ (see Appendix \ref{App3} for detailed evaluation) are given as
\begin{align}
&\mathcal{K}_1(p_1,p_2;m_1,\mu_1;m_2,\mu_2;m_3,\mu_3;\beta)
\nonumber\\&
=\frac{\left(\mu_1^2-m_1^2\right)-\left((\mu_2+p_1^0)^2-m_2^2\right)+\boldsymbol{p}_1^2}{2}
\mathcal{C}_{0}(p_1,p_2;m_1,\mu_1;m_2,\mu_2;m_3,\mu_3;\beta)
\nonumber\\&\hspace{0.4cm}
-\left(\mu_2-\mu_1\right)\mathscr{C}^{0}(p_1,p_2;m_1,\mu_1;m_2,\mu_2;m_3,\mu_3;\beta)
\nonumber\\&\hspace{0.4cm}
+\frac{\mathcal{B}_{0}(p_{1}+p_{2};m_1,\mu_1;m_3,\mu_3;\beta)-\mathcal{B}_{0}(p_{2};m_2,\mu_2;m_3,\mu_3;\beta)}{2},
\end{align}
\begin{align}
&\mathcal{K}_2(p_1,p_2;m_1,\mu_1;m_2,\mu_2;m_3,\mu_3;\beta)
\nonumber\\&
=\frac{\left[(\mu_2+p_1^0)^2-m_2^2-\boldsymbol{p}_1^2\right]
-\left[(\mu_3+p_1^0+p_2^0)^2-m_3^2-(\boldsymbol{p}_1+\boldsymbol{p}_2)^2\right]}{2}
\mathcal{C}_{0}(p_1,p_2;m_1,\mu_1;m_2,\mu_2;m_3,\mu_3;\beta)
\nonumber\\&\hspace{0.4cm}
-\left(\mu_3-\mu_2\right)\mathscr{C}^{0}(p_1,p_2;m_1,\mu_1;m_2,\mu_2;m_3,\mu_3;\beta)
\nonumber\\&\hspace{0.4cm}
+\frac{\mathcal{B}_{0}(p_{1};m_1,\mu_1;m_2,\mu_2;\beta)-\mathcal{B}_{0}(p_{1}+p_{2};m_1,\mu_1;m_3,\mu_3;\beta)}{2},
\end{align}
and
\begin{align}
\mathcal{K}_3(p_1,p_2;m_1,\mu_1;m_2,\mu_2;m_3,\mu_3;\beta)&=\mathscr{C}^{0}(p_1,p_2;m_1,\mu_1;m_2,\mu_2;m_3,\mu_3;\beta).
\end{align}
Obviously, $\mathcal{K}_{1}(p_1,p_2;m_1,\mu_1;m_2,\mu_2;m_3,\mu_3;\beta)$, $\mathcal{K}_{2}(p_1,p_2;m_1,\mu_1;m_2,\mu_2;m_3,\mu_3;\beta)$, and $\mathcal{K}_{3}(p_1,p_2;m_1,\mu_1;m_2,\mu_2;m_3,\mu_3;\beta)$ are functions of finite temperature and/or finite density, and they depend independently on $p_{1}^{0}$, $|\boldsymbol{p}_{1}|$, $p_{2}^{0}$ and $|\boldsymbol{p}_{2}|$, which are not Lorentz invariant. Consequently, the three-point generic one-loop tensor Feynman integral $\mathscr{C}^{\rho}(p_1,p_2;m_1,\mu_1;m_2,\mu_2;m_3,\mu_3;\beta)$ can be decomposed as linear combinations of two Lorentz-covariant tensors $p_{1}^{\rho}$ and $p_{2}^{\rho}$ and a non-Lorentz-covariant tensor $u^{\rho}$, with the non-Lorentz-invariant form factors $\mathcal{C}_{1}(p_1,p_2;m_1,\mu_1;m_2,\mu_2;m_3,\mu_3;\beta)$, $\mathcal{C}_{2}(p_1,p_2;m_1,\mu_1;m_2,\mu_2;m_3,\mu_3;\beta)$, and $\mathcal{C}_{3}(p_1,p_2;m_1,\mu_1;m_2,\mu_2;m_3,\mu_3;\beta)$ being expressed by a two-point generic one-loop scalar Feynman integral $\mathcal{B}_{0}(p_1;m_1,\mu_1;m_2,\mu_2;\beta)$, a three-point generic one-loop scalar Feynman integral $\mathcal{C}_{0}(p_1,p_2;m_1,\mu_1;m_2,\mu_2;m_3,\mu_3;\beta)$, and a temporal component of three-point generic one-loop tensor Feynman integral $\mathscr{C}^{0}(p_1,p_2;m_1,\mu_1;m_2,\mu_2;m_3,\mu_3;\beta)$, which are also non-Lorentz covariant.

By contrast, in the relativistic QFTs at zero temperature and zero density, the $D$-dimensional constant vector $u^{\rho}$ vanishes due to the Lorentz covariance, which leads to 
\begin{align}
&\mathscr{\tilde{C}}^{\rho}(p_1,p_2;m_1,\mu_1=0;m_2,\mu_2=0;m_3,\mu_3=0;\beta=\infty)
\nonumber\\&
=p_1^{\rho}\mathcal{\tilde{C}}_{1}(p_1,p_2;m_1,\mu_1=0;m_2,\mu_2=0;m_3,\mu_3=0;\beta=\infty)
\nonumber\\&\hspace{0.4cm}
+p_2^{\rho}\mathcal{\tilde{C}}_{2}(p_1,p_2;m_1,\mu_1=0;m_2,\mu_2=0;m_3,\mu_3=0;\beta=\infty)
\end{align}
in the \emph{conventional} Passarino-Veltman reduction, where the symbols with tilde $\mathcal{\tilde{C}}_{1}(p_1,p_2;m_1,\mu_1=0;m_2,\mu_2=0;m_3,\mu_3=0;\beta=\infty)$ and $\mathcal{\tilde{C}}_{2}(p_1,p_2;m_1,\mu_1=0;m_2,\mu_2=0;m_3,\mu_3=0;\beta=\infty)$ are introduced to denote the quantities at zero temperature. Contracting the three-point genetic one-loop tensor Feynman integral $\mathscr{\tilde{C}}^{\rho}(p_1,p_2;m_1,\mu_1=0;m_2,\mu_2=0;m_3,\mu_3=0;\beta=\infty)$ with $p_{1\rho}$ and $p_{2\rho}$ gives rise to two equations, 
\begin{align}
\tilde{G}_{C}^{(1)}(p_1,p_2)
\left(
  \begin{array}{c}
    \mathcal{\tilde{C}}_{1} \\\\
    \mathcal{\tilde{C}}_{2} \\
  \end{array}
\right)
=\left(
  \begin{array}{c}
    \mathcal{\tilde{K}}_{1} \\\\
    \mathcal{\tilde{K}}_{2} \\
  \end{array}
\right),
\label{COLTFI}
\end{align}
where $\tilde{G}_{C}^{(1)}(p_1,p_2)$ is the Gram matrix defined by
\begin{align}
\tilde{G}_{C}^{(1)}(p_1,p_2)
&=\left(
  \begin{array}{ccc}
    p_1^2 & p_1\cdot p_2 \\\\
    p_1\cdot p_2 & p_2^2 \\
  \end{array}
\right).
\end{align}

The arguments $(p_1,p_2;m_1,\mu_1=0;m_2,\mu_2=0;m_3,\mu_3=0;\beta=\infty)$ in $\mathcal{\tilde{C}}_{1}(p_1,p_2;m_1,\mu_1=0;m_2,\mu_2=0;m_3,\mu_3=0;\beta=\infty)$, $\mathcal{\tilde{C}}_{2}(p_1,p_2;m_1,\mu_1=0;m_2,\mu_2=0;m_3,\mu_3=0;\beta=\infty)$, $\mathcal{\tilde{K}}_{1}(p_1,p_2;m_1,\mu_1=0;m_2,\mu_2=0;m_3,\mu_3=0;\beta=\infty)$, and $\mathcal{\tilde{K}}_{2}(p_1,p_2;m_1,\mu_1=0;m_2,\mu_2=0;m_3,\mu_3=0;\beta=\infty)$ are omitted for short. The solutions to these two equations are given as
\begin{align}
\left(
  \begin{array}{c}
    \mathcal{\tilde{C}}_{1} \\\\\\
    \mathcal{\tilde{C}}_{2} \\
  \end{array}
\right)
=
\frac{1}{\tilde{\Delta}_{C}^{(1)}(p_1,p_2)}
\left(
  \begin{array}{cc}
  p_2^2 & -p_1\cdot p_2\\\\\\
  -p_1\cdot p_2  & p_1^2  \\
  \end{array}
\right)
\left(
  \begin{array}{c}
    \mathcal{\tilde{K}}_{1} \\\\\\
    \mathcal{\tilde{K}}_{2} \\
  \end{array}
\right),
\end{align}
where $\tilde{\Delta}_{C}^{(1)}(p_1,p_2)=p_1^2p_2^2-(p_1\cdot p_2)^{2}$ is the Gram determinant defined by the determinant of the Gram matrix $\tilde{G}_{C}^{(1)}(p_1,p_2)$. 

For the relativistic QFTs at zero temperature and zero density, it is evident that
\begin{align}
&\mathcal{\tilde{K}}_1(p_1,p_2;m_1,\mu_1=0;m_2,\mu_2=0;m_3,\mu_3=0;\beta=\infty)
\nonumber\\&
=\frac{m_2^2-m_1^2-p_1^2}{2}\mathcal{\tilde{C}}_{0}(p_1,p_2;m_1,\mu_1=0;m_2,\mu_2=0;m_3,\mu_3=0;\beta=\infty)
\nonumber\\&\hspace{0.4cm}
+\frac{\mathcal{\tilde{B}}_{0}(p_{1}+p_{2};m_1,\mu_1=0;m_3,\mu_3=0;\beta=\infty)
-\mathcal{\tilde{B}}_{0}(p_{2};m_2,\mu_2=0;m_3,\mu_3=0;\beta=\infty)}{2},\\
&\mathcal{\tilde{K}}_2(p_1,p_2;m_1,\mu_1=0;m_2,\mu_2=0;m_3,\mu_3=0;\beta=\infty)
\nonumber\\&
=\frac{m_3^2-m_2^2+p_1^2-(p_1+p_2)^2}{2}\mathcal{\tilde{C}}_{0}(p_1,p_2;\mu_1=0;m_2,\mu_2=0;m_3,\mu_3=0;\beta=\infty)
\nonumber\\&\hspace{0.4cm}
+\frac{\mathcal{\tilde{B}}_{0}(p_{1};m_1,\mu_1=0;m_2,\mu_2=0;\beta=\infty)-\mathcal{\tilde{B}}_{0}(p_{1}+p_{2};\mu_1=0;m_3,\mu_3=0;\beta=\infty)}{2},
\end{align}
which agree exactly with the results presented in the Appendix A of Ref.\cite{Ellis}. It is stressed that $\mathcal{\tilde{K}}_{1}(p_1,p_2;m_1,\mu_1=0;m_2,\mu_2=0;m_3,\mu_3=0;\beta=\infty)$ and $\mathcal{\tilde{K}}_{2}(p_1,p_2;m_1,\mu_1=0;m_2,\mu_2=0;m_3,\mu_3=0;\beta=\infty)$ can be obtained by directly setting $\mu_{1}=\mu_{2}=0$ and $T=0$ in Eqs. (\ref{F00Main}) and (\ref{F11Main}), respectively. It is evident that $\mathcal{\tilde{K}}_{1}(p_1,p_2;m_1,\mu_1=0;m_2,\mu_2=0;m_3,\mu_3=0;\beta=\infty)$ and $\mathcal{\tilde{K}}_{2}(p_1,p_2;m_1,\mu_1=0;m_2,\mu_2=0;m_3,\mu_3=0;\beta=\infty)$ are Lorentz invariant, and hence $\mathscr{\tilde{C}}^{\rho}(p_1,p_2;m_1,\mu_1=0;m_2,\mu_2=0;m_3,\mu_3=0;\beta=\infty)$ is Lorentz covariant.

\subsection{Reduction for three-point generic one-loop tensor Feynman integrals $\mathscr{C}^{\rho\sigma}(p_1,p_2;m_1,\mu_1;m_2,\mu_2;m_3,\mu_3;\beta)$ and $\mathscr{C}^{\rho\sigma\tau}(p_1,p_2;m_1,\mu_1;m_2,\mu_2;m_3,\mu_3;\beta)$}

The other two three-point genetic one-loop tensor Feynman integrals can be reduced in the similar way as
\begin{align}
&\mathscr{C}^{\rho\sigma}(p_1,p_2;m_1,\mu_1;m_2,\mu_2;m_3,\mu_3;\beta)
\nonumber\\&
=\int\frac{d^{D}l}{i\pi^{D/2}}\frac{l^{\rho}l^{\sigma}}
{\mathcal{P}(l,0;m_1,\mu_1)\mathcal{P}(l,p_1;m_2,\mu_2)\mathcal{P}(l,p_1+p_2;m_3,\mu_3)}
\nonumber\\&
=g^{\rho\sigma}\mathcal{C}_{00}(p_1,p_2;m_1,\mu_1;m_2,\mu_2;m_3,\mu_3;\beta)
\nonumber\\&\hspace{0.4cm}
+p_{1}^{\rho}p_{1}^{\sigma}\mathcal{C}_{11}(p_1,p_2;m_1,\mu_1;m_2,\mu_2;m_3,\mu_3;\beta)
\nonumber\\&\hspace{0.4cm}
+\left(p_{1}^{\rho}p_{2}^{\sigma}+p_{1}^{\sigma}p_{2}^{\rho}\right)\mathcal{C}_{12}(p_1,p_2;m_1,\mu_1;m_2,\mu_2;m_3,\mu_3;\beta)
\nonumber\\&\hspace{0.4cm}
+p_{2}^{\rho}p_{2}^{\sigma}\mathcal{C}_{22}(p_1,p_2;m_1,\mu_1;m_2,\mu_2;m_3,\mu_3;\beta)
\nonumber\\&\hspace{0.4cm}
+\left(p_{1}^{\rho}u^{\sigma}+p_{1}^{\sigma}u^{\rho}\right)\mathcal{C}_{13}(p_1,p_2;m_1,\mu_1;m_2,\mu_2;m_3,\mu_3;\beta)
\nonumber\\&\hspace{0.4cm}
+\left(p_{2}^{\rho}u^{\sigma}+p_{2}^{\sigma}u^{\rho}\right)\mathcal{C}_{23}(p_1,p_2;m_1,\mu_1;m_2,\mu_2;m_3,\mu_3;\beta)
\nonumber\\&\hspace{0.4cm}
+u^{\rho}u^{\sigma}\mathcal{C}_{33}(p_1,p_2;m_1,\mu_1;m_2,\mu_2;m_3,\mu_3;\beta),
\end{align}
and
\begin{align}
&\mathscr{C}^{\rho\sigma\tau}(p_1,p_2;m_1,\mu_1;m_2,\mu_2;m_3,\mu_3;\beta)
\nonumber\\&
=\int\frac{d^{D}l}{i\pi^{D/2}}\frac{l^{\rho}l^{\sigma}l^{\tau}}
{\mathcal{P}(l,0;m_1,\mu_1)\mathcal{P}(l,p_1;m_2,\mu_2)\mathcal{P}(l,p_1+p_2;m_3,\mu_3)}
\nonumber\\&
=\left(p_{1}^{\rho}g^{\sigma\tau}+p_{1}^{\sigma}g^{\tau\rho}+p_{1}^{\tau}g^{\rho\sigma}\right)
\mathcal{C}_{001}(p_1,p_2;m_1,\mu_1;m_2,\mu_2;m_3,\mu_3;\beta)
\nonumber\\&\hspace{0.4cm}
+\left(p_{2}^{\rho}g^{\sigma\tau}+p_{2}^{\sigma}g^{\tau\rho}+p_{2}^{\tau}g^{\rho\sigma}\right)
\mathcal{C}_{002}(p_1,p_2;m_1,\mu_1;m_2,\mu_2;m_3,\mu_3;\beta)
\nonumber\\&\hspace{0.4cm}
+\left(u^{\rho}g^{\sigma\tau}+u^{\sigma}g^{\tau\rho}+u^{\tau}g^{\rho\sigma}\right)
\mathcal{C}_{003}(p_1,p_2;m_1,\mu_1;m_2,\mu_2;m_3,\mu_3;\beta)
\nonumber\\&\hspace{0.4cm}
+p_{1}^{\rho}p_{1}^{\sigma}p_{1}^{\tau}\mathcal{C}_{111}(p_1,p_2;m_1,\mu_1;m_2,\mu_2;m_3,\mu_3;\beta)
\nonumber\\&\hspace{0.4cm}
+\left(p_{1}^{\rho}p_{1}^{\sigma}p_{2}^{\tau}+p_{1}^{\sigma}p_{1}^{\tau}p_{2}^{\rho}
+p_{1}^{\tau}p_{1}^{\rho}p_{2}^{\sigma}\right)
\mathcal{C}_{112}(p_1,p_2;m_1,\mu_1;m_2,\mu_2;m_3,\mu_3;\beta)
\nonumber\\&\hspace{0.4cm}
+\left(p_{1}^{\rho}p_{1}^{\sigma}u^{\tau}+p_{1}^{\sigma}p_{1}^{\tau}u^{\rho}
+p_{1}^{\tau}p_{1}^{\rho}u^{\sigma}\right)\mathcal{C}_{113}(p_1,p_2;m_1,\mu_1;m_2,\mu_2;m_3,\mu_3;\beta)
\nonumber\\&\hspace{0.4cm}
+\left(p_{1}^{\rho}p_{2}^{\sigma}p_{2}^{\tau}+p_{1}^{\sigma}p_{2}^{\tau}p_{2}^{\rho}
+p_{1}^{\tau}p_{2}^{\rho}p_{2}^{\sigma}\right)\mathcal{C}_{122}(p_1,p_2;m_1,\mu_1;m_2,\mu_2;m_3,\mu_3;\beta)
\nonumber\\&\hspace{0.4cm}
+p_{2}^{\rho}p_{2}^{\sigma}p_{2}^{\tau}\mathcal{C}_{222}(p_1,p_2;m_1,\mu_1;m_2,\mu_2;m_3,\mu_3;\beta)
\nonumber\\&\hspace{0.4cm}
+\left(p_{2}^{\rho}p_{2}^{\sigma}u^{\tau}+p_{2}^{\sigma}p_{2}^{\tau}u^{\rho}
+p_{2}^{\tau}p_{2}^{\rho}u^{\sigma}\right)\mathcal{C}_{223}(p_1,p_2;m_1,\mu_1;m_2,\mu_2;m_3,\mu_3;\beta)
\nonumber\\&\hspace{0.4cm}
+\left[p_{1}^{\rho}\left(p_{2}^{\sigma}u^{\tau}+p_{2}^{\tau}u^{\sigma}\right)
+p_{2}^{\rho}\left(p_{1}^{\sigma}u^{\tau}+p_{1}^{\tau}u^{\sigma}\right)
+u^{\rho}\left(p_{1}^{\sigma}p_{2}^{\tau}+p_{1}^{\tau}p_{2}^{\sigma}\right)\right]
\mathcal{C}_{123}(p_1,p_2;m_1,\mu_1;m_2,\mu_2;m_3,\mu_3;\beta)
\nonumber\\&\hspace{0.4cm}
+\left(p_{1}^{\rho}u^{\sigma}u^{\tau}+p_{1}^{\sigma}u^{\tau}u^{\rho}+p_{1}^{\tau}u^{\rho}u^{\sigma}\right)
\mathcal{C}_{133}(p_1,p_2;m_1,\mu_1;m_2,\mu_2;m_3,\mu_3;\beta)
\nonumber\\&\hspace{0.4cm}
+\left(p_{2}^{\rho}u^{\sigma}u^{\tau}+p_{2}^{\sigma}u^{\rho}u^{\tau}+p_{2}^{\tau}u^{\rho}u^{\sigma}\right)
\mathcal{C}_{233}(p_1,p_2;m_1,\mu_1;m_2,\mu_2;m_3,\mu_3;\beta)
\nonumber\\&\hspace{0.4cm}
+u^{\rho}u^{\sigma}u^{\tau}\mathcal{C}_{333}(p_1,p_2;m_1,\mu_1;m_2,\mu_2;m_3,\mu_3;\beta),
\end{align}
where the form factors satisfy $\mathcal{C}_{ab}=\mathcal{C}_{ba}$, $\mathcal{C}_{abc}=\mathcal{C}_{acb}=\mathcal{C}_{bac}=\mathcal{C}_{bca}=\mathcal{C}_{cab}=\mathcal{C}_{cba}$
and the arguments $(p_1,p_2;m_1,\mu_1;m_2,\mu_2;m_3,\mu_3;\beta)$ in them are omitted here for short. The form factors $\mathcal{C}_{ab}$ and $\mathcal{C}_{abc}$ can be expressed in terms of generic one-loop scalar Feynman integrals and one-loop tensor Feynman integrals up to three-point.

By contrast, for the relativistic QFTs at zero temperature and zero density, the $D$-dimensional constant vectors $u^{\rho}$, $u^{\sigma}$, and $u^{\tau}$ vanish due to the Lorentz covariance, which leads to
\begin{align}
&\mathscr{\tilde{C}}^{\rho\sigma}(p_1,p_2;m_1,\mu_1=0;m_2,\mu_2=0;m_3,\mu_3=0;\beta=\infty)
\nonumber\\&
=g^{\rho\sigma}\mathcal{\tilde{C}}_{00}(p_1,p_2;m_1,\mu_1=0;m_2,\mu_2=0;m_3,\mu_3=0;\beta=\infty)
\nonumber\\&\hspace{0.4cm}
+p_{1}^{\rho}p_{1}^{\sigma}\mathcal{\tilde{C}}_{11}(p_1,p_2;m_1,\mu_1=0;m_2,\mu_2=0;m_3,\mu_3=0;\beta=\infty)
\nonumber\\&\hspace{0.4cm}
+\left(p_{1}^{\rho}p_{2}^{\sigma}+p_{1}^{\sigma}p_{2}^{\rho}\right)
\mathcal{\tilde{C}}_{12}(p_1,p_2;m_1,\mu_1=0;m_2,\mu_2=0;m_3,\mu_3=0;\beta=\infty)
\nonumber\\&\hspace{0.4cm}
+p_{2}^{\rho}p_{2}^{\sigma}\mathcal{\tilde{C}}_{22}(p_1,p_2;m_1,\mu_1=0;m_2,\mu_2=0;m_3,\mu_3=0;\beta=\infty),
\end{align}
and
\begin{align}
&\mathscr{\tilde{C}}^{\rho\sigma\tau}(p_1,p_2;m_1,\mu_1=0;m_2,\mu_2=0;m_3,\mu_3=0;\beta=\infty)
\nonumber\\&
=\left(p_{1}^{\rho}g^{\sigma\tau}
+p_{1}^{\sigma}g^{\tau\rho}+p_{1}^{\tau}g^{\rho\sigma}\right)
\mathcal{\tilde{C}}_{001}(p_1,p_2;m_1,\mu_1=0;m_2,\mu_2=0;m_3,\mu_3=0;\beta=\infty)
\nonumber\\&\hspace{0.4cm}
+\left(p_{2}^{\rho}g^{\sigma\tau}
+p_{2}^{\sigma}g^{\tau\rho}+p_{2}^{\tau}g^{\rho\sigma}\right)
\mathcal{\tilde{C}}_{002}(p_1,p_2;m_1,\mu_1=0;m_2,\mu_2=0;m_3,\mu_3=0;\beta=\infty)
\nonumber\\&\hspace{0.4cm}
+p_{1}^{\rho}p_{1}^{\sigma}p_{1}^{\tau}\mathcal{\tilde{C}}_{111}(p_1,p_2;m_1,\mu_1=0;m_2,\mu_2=0;m_3,\mu_3=0;\beta=\infty)
\nonumber\\&\hspace{0.4cm}
+\left(p_{1}^{\rho}p_{1}^{\sigma}p_{2}^{\tau}+p_{1}^{\sigma}p_{1}^{\tau}p_{2}^{\rho}
+p_{1}^{\tau}p_{1}^{\rho}p_{2}^{\sigma}\right)\mathcal{\tilde{C}}_{112}(p_1,p_2;m_1,\mu_1=0;m_2,\mu_2=0;m_3,\mu_3=0;\beta=\infty)
\nonumber\\&\hspace{0.4cm}
+\left(p_{1}^{\rho}p_{2}^{\sigma}p_{2}^{\tau}+p_{1}^{\sigma}p_{2}^{\tau}p_{2}^{\rho}
+p_{1}^{\tau}p_{2}^{\rho}p_{2}^{\sigma}\right)\mathcal{\tilde{C}}_{122}(p_1,p_2;m_1,\mu_1=0;m_2,\mu_2=0;m_3,\mu_3=0;\beta=\infty)
\nonumber\\&\hspace{0.4cm}
+p_{2}^{\rho}p_{2}^{\sigma}p_{2}^{\tau}\mathcal{\tilde{C}}_{222}(p_1,p_2;m_1,\mu_1=0;m_2,\mu_2=0;m_3,\mu_3=0;\beta=\infty),
\end{align}
in the \emph{conventional} Passarino-Veltman reduction, where the symbols with tilde are introduced to denote the quantities at zero temperature.

\subsection{General tensor structures for the reduction of $N$-point generic one-loop tensor Feynman integrals}

The central step for the reduction of generic one-loop tensor Feynman integrals in the \emph{generalized} Passarino-Veltman reduction is to construct a complete set of tensor structures. For the $N$-point generic one-loop tensor Feynman integrals, we construct a complete set of tensor structures by utilizing the generic momenta $p$'s, the $D$-dimensional constant vectors $u$'s, and the metric tensors $g$'s. In a concise way, we use a notation in which curly braces denote symmetrization with respect to Lorentz indices \cite{CPVRS}. If there are only products of $m$ metric tensors $g$'s in the rank-$(2m)$ tensor structure, then we generally have
\begin{align}
\left\{g\cdots g\right\}^{\rho_{1}\rho_{2}\rho_{3}\rho_{4}\cdots\rho_{2m-1}\rho_{2m}}
&=\left\{g\right\}^{\rho_{1}\rho_{2}}\left\{g \cdots g\right\}^{\rho_{3}\rho_{4}\cdots\rho_{2m-1}\rho_{2m}}+\cdots\cdots,
\end{align}
where $``\cdots\cdots"$ denotes other nonequivalent permutations of the $2m$ Lorentz indices $(\rho_{1},\rho_{2},\rho_{3},\rho_{4},\cdots,\rho_{2m-1},\rho_{2m})$. Specifically, for $m=1$ and $m=2$, we have
\begin{align}
\left\{g\right\}^{\rho_{1}\rho_{2}}&=g^{\rho_{1}\rho_{2}},\\
\left\{g g\right\}^{\rho_{1}\rho_{2}\rho_{3}\rho_{4}}
&=\left\{g\right\}^{\rho_{1}\rho_{2}}\left\{g\right\}^{\rho_{3}\rho_{4}}
+\left\{g\right\}^{\rho_{1}\rho_{3}}\left\{g\right\}^{\rho_{2}\rho_{4}}
+\left\{g\right\}^{\rho_{1}\rho_{4}}\left\{g\right\}^{\rho_{2}\rho_{3}}.
\end{align}
If the rank-$m$ tensor structure consists purely of $m$ external momenta $p$'s, then we generally have
\begin{align}
\left\{p \cdots p\right\}_{i_{1}i_{2}i_{3}\cdots i_{m}}^{\rho_{1}\rho_{2}\rho_{3}\cdots \rho_{m-1}\rho_{m}}
&=p_{i_{1}}^{\rho_{1}}\left\{p \cdots p\right\}_{i_{2}i_{3}\cdots i_{m}}^{\rho_{2}\rho_{3}\cdots \rho_{m-1}\rho_{m}}+\cdots\cdots,
\end{align}
where $``\cdots\cdots"$ denotes other nonequivalent permutations of the $m$ Lorentz indices $(\rho_{1},\rho_{2},\cdots,\rho_{m-1},\rho_{m})$, and $(i_{1},i_{2},\cdots,i_{m-1},i_{m})$ label the momenta $(p_{i_{1}},p_{i_{2}},\cdots,p_{i_{m-1}},p_{i_{m}})$. Specifically, for $m=1$, $m=2$, and $m=3$, we have
\begin{align}
\left\{p\right\}_{i_{1}}^{\rho_{1}}&=p_{i_{1}}^{\rho_{1}},\\
\left\{p p\right\}_{i_{1}i_{2}}^{\rho_{1}\rho_{2}}
&=p_{i_{1}}^{\rho_{1}}p_{i_{2}}^{\rho_{2}}+p_{i_{1}}^{\rho_{2}}p_{i_{2}}^{\rho_{1}},\\
\left\{p p p\right\} _{i_{1}i_{2}i_{3}}^{\rho_{1}\rho_{2}\rho_{3}}
&=p_{i_{1}}^{\rho_{1}}\left\{p p\right\}_{i_{2}i_{3}}^{\rho_{2}\rho_{3}}
+p_{i_{1}}^{\rho_{2}}\left\{p p\right\}_{i_{2}i_{3}}^{\rho_{3}\rho_{1}}
+p_{i_{1}}^{\rho_{3}}\left\{p p\right\}_{i_{2}i_{3}}^{\rho_{1}\rho_{2}}.
\end{align}
When the rank-$m$ tensor structure is purely expressed by the product of $m$ constant vectors $u$'s, we generally have
\begin{align}
\left\{u \cdots u\right\}^{\rho_{1}\rho_{2}\cdots \rho_{m-1}\rho_{m}}
&=u^{\rho_{1}}u^{\rho_{2}}\cdots u^{\rho_{m-1}}u^{\rho_{m}}.
\end{align}
Specifically, for $m=1$ and $m=2$, we have
\begin{align}
\left\{u\right\}^{\rho_{1}}&=u^{\rho_{1}},\\
\left\{u u\right\}^{\rho_{1}\rho_{2}}&=u^{\rho_{1}}u^{\rho_{2}}.
\end{align}

If the rank-$(m+2)$ tensor structure is defined as the product of $m$ external momenta $p$'s and one metric tensor $g$, then we generally have
\begin{align}
\left\{p\cdots p g \right\}_{i_{1}i_{2}\cdots i_{m-1}i_{m}}^{\rho_{1}\rho_{2}\cdots\rho_{m-1}\rho_{m}\rho_{m+1}\rho_{m+2}}
&=\left\{p \cdots p\right\}_{i_{1}i_{2}\cdots i_{m-1}i_{m}}^{\rho_{1}\rho_{2}\cdots\rho_{m-1}\rho_{m}}g^{\rho_{m+1}\rho_{m+2}}
+\cdots\cdots,
\end{align}
where $``\cdots\cdots"$ denotes other nonequivalent permutations of the $(m$+$2)$ Lorentz indices $(\rho_{1},\rho_{2},\cdots,\rho_{m-1},\rho_{m},\rho_{m+1},\rho_{m+2})$, and $(i_{1},i_{2},\cdots,i_{m-1},i_{m})$ label the $m$ momenta $(p_{i_{1}},p_{i_{2}},\cdots,p_{i_{m-1}},p_{i_{m}})$. Specifically, for $m=1$ and $m=2$, we have
\begin{align}
\left\{p g\right\}_{i_{1}}^{\rho_{1}\rho_{2}\rho_{3}}
&=p_{i_{1}}^{\rho_{1}}\left\{g\right\}^{\rho_{2}\rho_{3}}+p_{i_{1}}^{\rho_{2}}\left\{g\right\}^{\rho_{3}\rho_{1}}
+p_{i_{1}}^{\rho_{3}}\left\{g\right\}^{\rho_{1}\rho_{2}},\\
\left\{p p g\right\}_{i_{1}i_{2}}^{\rho_{1}\rho_{2}\rho_{3}\rho_{4}}
&=\left\{pp\right\}_{i_{1}i_{2}}^{\rho_{1}\rho_{2}}\left\{g\right\}^{\rho_{3}\rho_{4}}
+\left\{pp\right\}_{i_{1}i_{2}}^{\rho_{1}\rho_{3}}\left\{g\right\}^{\rho_{2}\rho_{4}}
+\left\{pp\right\}_{i_{1}i_{2}}^{\rho_{1}\rho_{4}}\left\{g\right\}^{\rho_{2}\rho_{3}}
\nonumber\\&
+\left\{pp\right\}_{i_{1}i_{2}}^{\rho_{2}\rho_{3}}\left\{g\right\}^{\rho_{4}\rho_{1}}
+\left\{pp\right\}_{i_{1}i_{2}}^{\rho_{2}\rho_{4}}\left\{g\right\}^{\rho_{1}\rho_{3}}
+\left\{pp\right\}_{i_{1}i_{2}}^{\rho_{3}\rho_{4}}\left\{g\right\}^{\rho_{1}\rho_{2}}.
\end{align}

If the rank-$(m+2)$ tensor structure consists of $m$ constant vectors $u$ and one metric tensor $g$, then we generally have
\begin{align}
\left\{u\cdots u g \right\}^{\rho_{1}\cdots\rho_{m}\rho_{m+1}\rho_{m+2}}
&=\left\{u\cdots u\right\}^{\rho_{1}\cdots\rho_{m}}\left\{g\right\}^{\rho_{m+1}\rho_{m+2}}
+\cdots\cdots,
\end{align}
where $``\cdots\cdots"$ denotes other nonequivalent permutations of the $(m$+$2)$ Lorentz indices $(\rho_{1},\rho_{2},\cdots,\rho_{m-1},\rho_{m},\rho_{m+1},\rho_{m+2})$. Specifically, for $m=1$ and $m=2$, we have
\begin{align}
\left\{u g\right\}^{\rho_{1}\rho_{2}\rho_{3}}
&=u^{\rho_{1}}\left\{g\right\}^{\rho_{2}\rho_{3}}+u^{\rho_{2}}\left\{g\right\}^{\rho_{3}\rho_{1}}
+u^{\rho_{3}}\left\{g\right\}^{\rho_{1}\rho_{2}},\\
\left\{u u g \right\}^{\rho_{1}\rho_{2}\rho_{3}\rho_{4}}
&=\left\{u u\right\}^{\rho_{1}\rho_{2}}\left\{g\right\}^{\rho_{3}\rho_{4}}
+\left\{u u\right\}^{\rho_{1}\rho_{3}}\left\{g\right\}^{\rho_{2}\rho_{4}}
+\left\{u u\right\}^{\rho_{1}\rho_{4}}\left\{g\right\}^{\rho_{2}\rho_{3}}
\nonumber\\&
+\left\{u u\right\}^{\rho_{2}\rho_{3}}\left\{g\right\}^{\rho_{1}\rho_{4}}
+\left\{u u\right\}^{\rho_{2}\rho_{4}}\left\{g\right\}^{\rho_{1}\rho_{3}}
+\left\{u u\right\}^{\rho_{3}\rho_{4}}\left\{g\right\}^{\rho_{1}\rho_{2}}.
\end{align}

If there are $m$ external momenta $p$'s and $n$ constant vectors $u$'s in the rank-$(m+n)$ tensor structure, then we generally have
\begin{align}
\left\{p\cdots p u\cdots u\right\}_{i_{1}\cdots i_{m}}^{\rho_{1}\cdots \rho_{m}\rho_{m+1}\cdots\rho_{m+n}}
&=\left\{p\cdots p \right\}_{i_{1}\cdots i_{m}}^{\rho_{1}\cdots \rho_{m}}\left\{u\cdots u \right\}^{\rho_{m+1}\cdots \rho_{m+n}}
+\cdots\cdots,
\end{align}
where $``\cdots\cdots"$ denotes other nonequivalent permutations of the $(m$+$2)$ Lorentz indices $(\rho_{1},\cdots,\rho_{m-1},\rho_{m},\cdots,\rho_{m+n})$, and $(i_{1},i_{2},\cdots,i_{m-1},i_{m})$ label the $m$ momenta $(p_{i_{1}},p_{i_{2}},\cdots,p_{i_{m-1}},p_{i_{m}})$. Specifically, we have
\begin{align}
&\left\{p u\right\}_{i_{1}}^{\rho_{1}\rho_{2}}
=p_{i_{1}}^{\rho_{1}}u^{\rho_{2}}+u^{\rho_{1}}p_{i_{1}}^{\rho_{2}},\\
&\left\{p p u\right\}_{i_{1}i_{2}}^{\rho_{1}\rho_{2}\rho_{3}}
=\left\{p p\right\}_{i_{1}i_{2}}^{\rho_{1}\rho_{2}}u^{\rho_{3}}
+\left\{p p\right\}_{i_{1}i_{2}}^{\rho_{2}\rho_{3}}u^{\rho_{1}}
+\left\{p p\right\}_{i_{1}i_{2}}^{\rho_{3}\rho_{1}}u^{\rho_{2}},\\
&\left\{p u u\right\}_{i_{1}}^{\rho_{1}\rho_{2}\rho_{3}}
=p_{i_{1}}^{\rho_{1}}\left\{u u\right\}^{\rho_{2}\rho_{3}}
+p_{i_{1}}^{\rho_{2}}\left\{u u\right\}^{\rho_{3}\rho_{1}}
+p_{i_{1}}^{\rho_{3}}\left\{u u\right\}^{\rho_{1}\rho_{2}}.
\end{align}

For the most general case in which the rank-$(m+n+2l)$ tensor structure is defined as the product of $m$ external momenta $p$'s, $n$ constant vectors $u$'s, and $l$ metric tensors $g$'s, we have
\begin{align}
&\left\{p\cdots p u\cdots u g\cdots g \right\}_{i_{1}\cdots i_{m}}^{\rho_{1}\cdots \rho_{m}\rho_{m+1}\cdots \rho_{m+n}\rho_{m+n+1}\cdots \rho_{m+n+2l}}
\nonumber\\
&=\left\{p\cdots p\right\}_{i_{1}\cdots i_{m}}^{\rho_{1}\cdots\rho_{m}}\left\{u\cdots u\right\}^{\rho_{m+1}\cdots\rho_{m+n}}
\left\{g\cdots g\right\}^{\rho_{m+n+1}\cdots \rho_{m+n+2l}}
+\cdots\cdots,
\end{align}
where $``\cdots\cdots"$ denotes other nonequivalent permutations of the $(m+n+2l)$ Lorentz indices $(\rho_{1},\cdots, \rho_{m},\rho_{m+1},\cdots, \rho_{m+n},\rho_{m+n+1},\cdots, \rho_{m+n+2l})$, and $(i_{1},i_{2},\cdots,i_{m-1},i_{m})$ label the $m$ momenta $(p_{i_{1}},p_{i_{2}},\cdots,p_{i_{m-1}},p_{i_{m}})$. Specifically, for $m=1$, $n=1$, and $l=1$, we have
\begin{align}
&\left\{p u g\right\}_{i_{1}}^{\rho_{1}\rho_{2}\rho_{3}\rho_{4}}
=g^{\rho_{1}\rho_{2}}\left\{p u\right\}_{i_{1}}^{\rho_{3}\rho_{4}}
+g^{\rho_{2}\rho_{3}}\left\{p u\right\}_{i_{1}}^{\rho_{1}\rho_{4}}
+g^{\rho_{3}\rho_{1}}\left\{p u\right\}_{i_{1}}^{\rho_{2}\rho_{4}}.
\end{align}

Obviously, $p_{i_{a}}^{\rho}$ and $u^{\rho}$ with $a=1,2,\cdots$, $N-1$ form a complete set of rank-one tensor structures for the rank-one $N$-point generic one-loop tensor Feynman integrals $\mathscr{I}^{\rho}(p_1,\cdots,p_{N-1};m_1,\mu_1;\cdots;m_N,\mu_N;\beta)$. Specifically, $p_{1}^{\rho}$ and $u^{\rho}$ form a complete set of rank-one tensor structures for the rank-one two-point ($N=2$) generic one-loop tensor Feynman integrals $\mathscr{B}^{\rho}(p_1;m_1,\mu_1;m_2,\mu_2;\beta)$. Similarly, $p_{1}^{\rho}$, $p_{2}^{\rho}$, and $u^{\rho}$ form a complete set of rank-one tensor structures for the rank-one three-point ($N=3$) generic one-loop tensor Feynman integrals $\mathscr{C}^{\rho}(p_1,p_2;m_1,\mu_1;m_2,\mu_2;m_3,\mu_3;\beta)$.

In addition, $\left\{g\right\}^{\rho\sigma}$, $\left\{p p\right\}_{i_{a}i_{b}}^{\rho\sigma}$, $\left\{p u\right\}_{i_{a}}^{\rho\sigma}$, and $\left\{u u\right\}^{\rho\sigma}$ with $a,b=1,2,\cdots,N-1$ form a complete set of rank-two tensor structures for the rank-two $N$-point generic one-loop tensor Feynman integrals $\mathscr{I}^{\rho\sigma}(p_1,\cdots,p_{N-1};m_1,\mu_1;\cdots;m_N,\mu_N;\beta)$. Specifically, the complete set of rank-two tensor structures for the rank-two two-point generic one-loop tensor Feynman integrals $\mathscr{B}^{\rho\sigma}(p_1;m_1,\mu_1;m_2,\mu_2;\beta)$ can be constructed by
\begin{align}
\left\{g\right\}^{\rho\sigma}&=g^{\rho\sigma},\\
\left\{p p\right\}_{11}^{\rho\sigma}&=p_{1}^{\rho}p_{1}^{\sigma},\\
\left\{p u\right\}_{1}^{\rho\sigma}&=p_{1}^{\rho}u^{\sigma}+p_{1}^{\sigma}u^{\rho},\\
\left\{u u\right\}^{\rho\sigma}&=u^{\rho}u^{\sigma}.
\end{align}
Similarly, the complete set of rank-two tensor structures for the rank-two three-point generic one-loop tensor Feynman integrals $\mathscr{C}^{\rho\sigma}(p_1,m_1,\mu_1;m_2,\mu_2;m_3,\mu_3;\beta)$, can be constructed by
\begin{align}
\left\{g\right\}^{\rho\sigma}&=g^{\rho\sigma},\\
\left\{p p\right\}_{11}^{\rho\sigma}&=p_{1}^{\rho}p_{1}^{\sigma},\\
\left\{p p\right\}_{12}^{\rho\sigma}&=p_{1}^{\rho}p_{2}^{\sigma}+p_{1}^{\sigma}p_{2}^{\rho},\\
\left\{p p\right\}_{22}^{\rho\sigma}&=p_{2}^{\rho}p_{2}^{\sigma},\\
\left\{p u\right\}_{1}^{\rho\sigma}&=p_{1}^{\rho}u^{\sigma}+p_{1}^{\sigma}u^{\rho},\\
\left\{p u\right\}_{2}^{\rho\sigma}&=p_{2}^{\rho}u^{\sigma}+p_{2}^{\sigma}u^{\rho},\\
\left\{u u\right\}^{\rho\sigma}&=u^{\rho}u^{\sigma}.
\end{align}

The parallel procedure can be applied to the reduction of rank-three generic one-loop tensor Feynman integrals. The rank-three tensor structures $\left\{p g\right\}_{i_{a}}^{\rho\sigma\tau}$, $\left\{u g\right\}^{\rho\sigma\tau}$, $\left\{p p p\right\}_{i_{a}i_{b}i_{c}}^{\rho\sigma\tau}$, $\left\{p p u\right\}_{i_{a}i_{b}}^{\rho\sigma\tau}$, $\left\{p u u\right\}_{i_{a}}^{\rho\sigma\tau}$, and $\left\{u u u\right\}^{\rho\sigma\tau}$, with $a,b,c=1,2,\cdots,N-1$ form a complete set for the rank-three $N$-point generic one-loop tensor Feynman integrals $\mathscr{I}^{\rho\sigma\tau}(p_1,\cdots,p_{N-1};m_1,\mu_1;\cdots;m_N,\mu_N;\beta)$. Namely, the complete set of rank-three tensor structures for the rank-three three-point generic one-loop tensor Feynman integrals $\mathscr{C}^{\rho\sigma\tau}(p_1,p_2;m_1,\mu_1;m_2,\mu_2;m_3,\mu_3;\beta)$ can be constructed by
\begin{align}
\left\{p g\right\}_{1}^{\rho\sigma\tau}&=p_{1}^{\rho}g^{\sigma\tau}+p_{1}^{\sigma}g^{\tau\rho}+p_{1}^{\tau}g^{\rho\sigma},\\
\left\{p g\right\}_{2}^{\rho\sigma\tau}&=p_{2}^{\rho}g^{\sigma\tau}+p_{2}^{\sigma}g^{\tau\rho}+p_{2}^{\tau}g^{\rho\sigma},\\
\left\{u g\right\}^{\rho\sigma\tau}&=u^{\rho}g^{\sigma\tau}+u^{\sigma}g^{\tau\rho}+u^{\tau}g^{\rho\sigma},\\
\left\{p p p\right\}_{111}^{\rho\sigma\tau}&=p_{1}^{\rho}p_{1}^{\sigma}p_{1}^{\tau},\\
\left\{p p p\right\}_{112}^{\rho\sigma\tau}&=p_{1}^{\rho}p_{1}^{\sigma}p_{2}^{\tau}+p_{1}^{\sigma}p_{1}^{\tau}p_{2}^{\rho}
+p_{1}^{\tau}p_{1}^{\rho}p_{2}^{\sigma},\\
\left\{p p u\right\}_{11}^{\rho\sigma\tau}&=p_{1}^{\rho}p_{1}^{\sigma}u^{\tau}+p_{1}^{\sigma}p_{1}^{\tau}u^{\rho}
+p_{1}^{\tau}p_{1}^{\rho}u^{\sigma},\\
\left\{p p p\right\}_{122}^{\rho\sigma\tau}&=p_{1}^{\rho}p_{2}^{\sigma}p_{2}^{\tau}+p_{1}^{\sigma}p_{2}^{\tau}p_{2}^{\rho}
+p_{1}^{\tau}p_{2}^{\rho}p_{2}^{\sigma},\\
\left\{p p p\right\}_{222}^{\rho\sigma\tau}&=p_{2}^{\rho}p_{2}^{\sigma}p_{2}^{\tau},\\
\left\{p p u\right\}_{22}^{\rho\sigma\tau}&=p_{2}^{\rho}p_{2}^{\sigma}u^{\tau}+p_{2}^{\sigma}p_{2}^{\tau}u^{\rho}
+p_{2}^{\tau}p_{2}^{\rho}u^{\sigma},\\
\left\{p p p\right\}_{12}^{\rho\sigma\tau}&=p_{1}^{\rho}\left(p_{2}^{\sigma}u^{\tau}+p_{2}^{\tau}u^{\sigma}\right)
+p_{2}^{\rho}\left(p_{1}^{\sigma}u^{\tau}+p_{1}^{\tau}u^{\sigma}\right)
+u^{\rho}\left(p_{1}^{\sigma}p_{2}^{\tau}+p_{1}^{\tau}p_{2}^{\sigma}\right),\\
\left\{p u u\right\}_{1}^{\rho\sigma\tau}&=p_{1}^{\rho}u^{\sigma}u^{\tau}
+p_{1}^{\sigma}u^{\tau}u^{\rho}+p_{1}^{\tau}u^{\rho}u^{\sigma},\\
\left\{p u u\right\}_{2}^{\rho\sigma\tau}&=p_{2}^{\rho}u^{\sigma}u^{\tau}
+p_{2}^{\sigma}u^{\tau}u^{\rho}+p_{2}^{\tau}u^{\rho}u^{\sigma},\\
\left\{u u u\right\}^{\rho\sigma\tau}&=u^{\rho}u^{\sigma}u^{\tau}.
\end{align}

We end this section by emphasizing the essentials of \emph{generalized} Passarino-Veltman reduction with that of \emph{conventional} Passarino-Veltman reduction due to the explicit breaking of Lorentz covariance in the relativistic QFTs at finite temperature and/or finite density. First, besides the Lorentz-covariant tensor structures appeared in the \emph{conventional} Passarino-Veltman reduction \cite{TVSI,CPVRS}, several other non-Lorentz-covariant tensor structures where the non-Lorentz-covariant $D$-dimensional constant vector $u^{\rho}$ which appears at least once are needed in the \emph{generalized} Passarino-Veltman reduction. Second, up to $N$-point one-loop Feynman diagrams, besides the generic one-loop scalar Feynman integrals $\mathcal{I}_{n}$ with $1\le n\le N$ which are complete to expand the generic one-loop tensor Feynman integrals in the \emph{conventional} Passarino-Veltman reduction \cite{TVSI,CPVRS}, several purely temporal components of generic one-loop tensor Feynman integrals, $\mathscr{I}_{n}^{0}
=u_{\rho}\mathscr{I}_{n}^{\rho}$, $\mathscr{I}_{n}^{00}
=u_{\rho}u_{\sigma}\mathscr{I}_{n}^{\rho\sigma}$, $\mathscr{I}_{n}^{000}
=u_{\rho}u_{\sigma}u_{\tau}\mathscr{I}_{n}^{\rho\sigma\tau}$ and so forth with $1\le n\le N$, are also needed in the \emph{generalized} Passarino-Veltman reduction to form a complete set of elementary one-loop Feynman integrals for expanding the generic one-loop tensor Feynman integrals $\mathscr{I}_{N}^{\rho}$, $\mathscr{I}_{N}^{\rho\sigma}$, $\mathscr{I}_{N}^{\rho\sigma\tau}$, and so forth. Third, in contrast to the Lorentz-invariant counterparts depending on $p_{n}^{2}=p_{n}\cdot p_{n}$ in the \emph{conventional} Passarino-Veltman reduction \cite{TVSI,CPVRS}, the generic one-loop scalar Feynman integrals in the \emph{generalized} Passarino-Veltman reduction depend independently on $p_{i_{a}}^{0}=u\cdot p_{i_{a}}$ and $|\boldsymbol{p}_{i_{a}}|=\sqrt{(u\cdot p_{i_{a}})^{2}-p_{i_{a}}^{2}}$ with $a=1,2,\cdots,N-1$, which are no longer Lorentz-invariant \cite{Rehberg1996AOP,Rehberg1996PRC}. In addition, the purely temporal components of generic one-loop tensor Feynman integrals are also non-Lorentz invariant. Fourth, for the relativistic QFTs at zero temperature and zero density, the Lorentz covariance is restored for the absence of preferred rest reference frame. Therefore, the $D$-dimensional constant vector  is not defined and always vanishes. As a consequence, one forces $u^{\rho}$ to be zero in the expressions of all the one-loop tensor Feynman integrals in terms of tensor structures for the relativistic QFTs at zero temperature and zero density, and hence the \emph{generalized} Passarino-Veltman reduction goes back to the \emph{conventional} Passarino-Veltman reduction.

\section{Demonstration applications\label{Applications}}

In this section, we present demonstration applications of \emph{generalized} Passarino-Veltman reduction for simplifying the one-loop pseudoscalar polarization function in the Nambu-Jona-Lasinio (NJL) model and the one-loop photon self-energy in the $D$-dimensional quantum electrodynamics (QED), respectively.

\subsection{One-loop pseudoscalar polarization function in the NJL model}

As the first demonstration application, we utilize the \emph{generalized} Passarino-Veltman reduction to calculate the one-loop pseudoscalar polarization function at finite temperature and/or finite density in the NJL model \cite{RMP1992NJL,PhysRep1994NJL}. In the three-flavor version, the Lagrangian takes \cite{Rehberg1996PRC,Rehberg1996AOP}
\begin{align}
\mathscr{L}_{\mathrm{NJL}}&=\sum_{f=u,d,s}\bar{\psi}_{f}\left[i\gamma^\rho\partial_\rho-m_{f}+\gamma^0\mu_{f}\right]\psi_{f}
+G\sum_{a=0}^{8}\left[\left(\bar{\psi}\lambda^{a}\psi\right)^{2}+\left(\bar{\psi}i\gamma^{5}\lambda^{a}\psi\right)^{2}\right]
\nonumber\\&
-K\left[\det\bar{\psi}\left(1+\gamma^{5}\right)\psi+\det\bar{\psi}\left(1-\gamma^{5}\right)\psi\right],
\label{LNJL}
\end{align}
where we have added the term related to the chemical potential $\mu_{f}$ for a quark of flavor $f$. In this expression, $G$ and $K$ are dimensionful coupling constants, $\lambda^{a}$ denotes the Gell-Mann matrices in flavor space, and $m_{f}$ is the flavor-dependent current quark mass, respectively. The flavor index $f$ carried by the quark fields $\psi_{f}$ is only shown in the free part of the Lagrangian, and the color index is generally suppressed.

\begin{figure}[htbp]
\centering
\includegraphics[width=5cm]{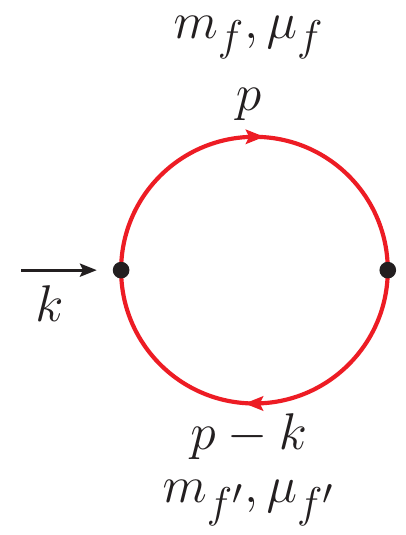}
\caption{One-loop pseudoscalar polarization function $-i\Pi_{f\bar{f}^{\prime}}^{\mathrm{PS}}(k;m_{f},\mu_{f};m_{f^{\prime}},\mu_{f^{\prime}};\beta)$ in the NJL model.}
\label{FigDiagNJL}
\end{figure}

Following the notations in the Refs. \cite{Rehberg1996PRC,Rehberg1996AOP}, we write down the one-loop pseudoscalar polarization function in the NJL model [see Fig. \ref{FigDiagNJL}] as
\begin{align}
&-i\Pi_{f\bar{f}^{\prime}}^{\mathrm{PS}}(k;m_{f},\mu_{f};m_{f^{\prime}},\mu_{f^{\prime}};\beta)
\nonumber\\&
=-N_{c}\int\frac{d^{4}l}{(2\pi)^{4}}\frac{\mathrm{tr}\left\{\gamma^{5}\left[l^\rho\gamma_\rho+\mu_{f}\gamma_0+m_{f}\right]
\gamma^{5}\left[(l^\sigma-k^\sigma)\gamma_\sigma+\mu_{f^{\prime}}\gamma_0+m_{f^{\prime}}\right]\right\}}{
\left[(l^0+\mu_{f})^2-\boldsymbol{l}^2-m_{f}^2\right]
\left[(l^0+k^0+\mu_{f^{\prime}})^2-(\boldsymbol{l}-\boldsymbol{k})^2-m_{f^{\prime}}^2\right]}
\nonumber\\&
=4 N_{c}\int\frac{d^{4}l}{(2\pi)^{4}}\frac{g_{\rho\sigma}l^{\rho}(l^{\sigma}-k^{\sigma})+g_{\rho 0}l^{\rho}(\mu_{f}+\mu_{f^{\prime}})
-g_{\rho 0}k^{\rho}\mu_{f}+(\mu_{f}\mu_{f^{\prime}}-m_{f}m_{f^{\prime}})}{
\left[(l^0+\mu_{f})^2-\boldsymbol{l}^2-m_{f}^2\right]
\left[(l^0+q^0+\mu_{f^{\prime}})^2-(\boldsymbol{l}+\boldsymbol{q})^2-m_{f^{\prime}}^2\right]}.
\end{align}
In terms of two generic one-loop tensor Feynman integrals $\mathscr{B}^{\rho}(k;m_{f},\mu_{f};m_{f^{\prime}},\mu_{f^{\prime}};\beta)$ and $\mathscr{B}^{\rho\sigma}(k;m_{f},\mu_{f};m_{f^{\prime}},\mu_{f^{\prime}};\beta)$, and a generic one-loop scalar Feynman integral $\mathcal{B}_{0}(k;m_{f},\mu_{f};m_{f^{\prime}},\mu_{f^{\prime}};\beta)$, this one-loop pseudoscalar polarization function can be decomposed into
\begin{align}
&-i\Pi_{f\bar{f}^{\prime}}^{\mathrm{PS}}(k;m_{f},\mu_{f};m_{f^{\prime}},\mu_{f^{\prime}};\beta)
\nonumber\\&
=\frac{4iN_{c}}{(4\pi)^{2}}
\left\{g_{\rho\sigma}\Big[\mathscr{B}^{\rho\sigma}\left(k;m_{f},\mu_{f};m_{f^{\prime}},\mu_{f^{\prime}};\beta\right)
-\mathscr{B}^{\rho}\left(k;m_{f},\mu_{f};m_{f^{\prime}},\mu_{f^{\prime}}\right)k^{\sigma}\Big]
\right.\nonumber\\&\hspace{1.6cm}
+g_{\rho 0}\mathscr{B}^{\rho}\left(k;m_{f},\mu_{f};m_{f^{\prime}},\mu_{f^{\prime}};\beta\right)\left(\mu_{f}+\mu_{f^{\prime}}\right)
-g_{\rho 0}k^{\rho}\mathcal{B}_{0}\left(k;m_{f},\mu_{f};m_{f^{\prime}},\mu_{f^{\prime}};\beta\right)
\nonumber\\&\hspace{1.6cm}\left.
+\left(\mu_{f}\mu_{f^{\prime}}-m_{f}m_{f^{\prime}}\right)
\mathcal{B}_{0}\left(k;m_{f},\mu_{f};m_{f^{\prime}},\mu_{f^{\prime}};\beta\right)\Big.\right\}.
\label{PSP}
\end{align}
After inserting $\mathscr{B}^{\rho}(p_{1};m_{1},\mu_{1};m_{2},\mu_{2};\beta)$ and $\mathscr{B}^{\rho\sigma}(p_{1};m_{1},\mu_{1};m_{2},\mu_{2};\beta)$ in Sec.  \ref{ROLTFI} into Eq. (\ref{PSP}) by setting $\mu_1=\mu_{f}$, $\mu_2=\mu_{f^{\prime}}$, $m_1=m_{f}$, $m_2=m_{f^{\prime}}$, and $p_{1}=k$, the one-loop pseudoscalar polarization function can be further written as
\begin{align}
&-i\Pi_{f\bar{f}^{\prime}}^{\mathrm{PS}}(k;m_{f},\mu_{f};m_{f^{\prime}},\mu_{f^{\prime}};\beta)
\nonumber\\&
=\frac{iN_{c}}{8\pi^{2}}
\left\{\Big.\mathcal{A}_{0}\left(0;m_{f},\mu_{f};\beta\right)+\mathcal{A}_{0}\left(0;m_{f^{\prime}},\mu_{f^{\prime}};\beta\right)
\right.\nonumber\\&\hspace{1.2cm}\left.
+\Big[\left(m_{f}-m_{f^{\prime}}\right)^2
-\left(k^{0}+\mu_{f}-\mu_{f^{\prime}}\right)^{2}-\boldsymbol{k}^2\Big]
\mathcal{B}_{0}\left(k;m_{f},\mu_{f};m_{f^{\prime}},\mu_{f^{\prime}};\beta\right)\right\}.
\end{align}

Interestingly, it is adequate to express the one-loop pseudoscalar polarization function by utilizing the one-loop scalar Feynman integrals $\mathcal{A}_{0}(0;m_{f},\mu_{f};\beta)$, $\mathcal{A}_{0}(0;m_{f^{\prime}},\mu_{f^{\prime}};\beta)$, and $\mathcal{B}_{0}(k;m_{f},\mu_{f};m_{f^{\prime}},\mu_{f^{\prime}};\beta)$, without resorting to the temporal components of one-loop tensor Feynman integrals $\mathscr{B}^{0}(k;m_{f},\mu_{f};m_{f^{\prime}},\mu_{f^{\prime}};\beta)$ and $\mathscr{B}^{00}(k;m_{f},\mu_{f};m_{f^{\prime}},\mu_{f^{\prime}};\beta)$. It is emphasized that the one-loop pseudoscalar polarization function here agrees exactly with the previous results \cite{Rehberg1996PRC,Rehberg1996AOP}, which can be taken as a benchmark to test the \emph{generalized} Passarino-Veltman reduction.

\subsection{One-loop photon self-energy in the $D$-dimensional QED}

As the second demonstration application, we employ the \emph{generalized} Passarino-Veltman reduction to perform the one-loop photon self-energy at finite temperature and/or finite density in the $D$-dimensional QED, whose Lagrangian in the Feynman gauge reads
\begin{align}
\mathscr{L}_{\mathrm{QED}}&=\bar{\psi}\left[i\gamma^\rho(\partial_\rho+ieA_\rho)-m+\gamma^0\mu\right]\psi
-\frac{1}{4}F_{\rho\sigma}F^{\rho\sigma}-\frac{1}{2}\left(\partial^{\rho}A_{\rho}\right)^2,
\label{LQED}
\end{align}
where $e$, $m$, and $\mu$ denote the charge, mass, and chemical potential of electrons, respectively. 

\begin{figure}[htbp]
\centering
\includegraphics[width=5cm]{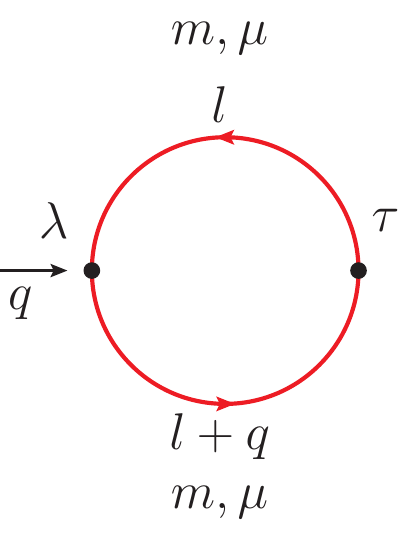}
\caption{One-loop photon self-energy $i\Pi^{\lambda\tau}(q;m,\mu;m,\mu;\beta)$ in the $D$-dimensional QED.}
\label{FigDiagQED}
\end{figure}

The one-loop photon self-energy in the $D$-dimensional QED [see Fig. \ref{FigDiagQED}] can be expressed as
\begin{align}
&i\Pi^{\lambda\tau}(q;m,\mu;m,\mu;\beta)
\nonumber\\&
=-e^{2}\int\frac{d^Dl}{(2\pi)^{D}}\frac{\mathrm{tr}\left\{\gamma^\lambda\left[l^\rho\gamma_\rho+\mu\gamma_{0}+m\right]
\gamma^\tau\left[(l^\sigma+q^\sigma)\gamma_\sigma+\mu\gamma_{0}+m\right]\right\}}{
\left[(l^0+\mu)^2-\boldsymbol{l}^2-m^2\right]
\left[(l^0+q^0+\mu)^2-(\boldsymbol{l}+\boldsymbol{q})^2-m^2\right]}
\nonumber\\&
=-4e^{2}\int\frac{d^Dl}{(2\pi)^{D}}\frac{
\left[g_{~\rho}^{\lambda}g_{~\sigma}^{\tau}-g^{\lambda\tau}g_{\rho\sigma}+g_{~\sigma}^{\lambda}g_{~\rho}^{\tau}\right]
l^{\rho}(l^{\sigma}+q^{\sigma})}{
\left[(l^0+\mu)^2-\boldsymbol{l}^2-m^2\right]
\left[(l^0+q^0+\mu)^2-(\boldsymbol{l}+\boldsymbol{q})^2-m^2\right]}
\nonumber\\&\hspace{0.3cm}
-4e^{2}\int\frac{d^Dl}{(2\pi)^{D}}\frac{
\left[g_{~\rho}^{\lambda}g_{~0}^{\tau}-g^{\lambda\tau}g_{\rho0}+g_{~0}^{\lambda}g_{~\rho}^{\tau}\right]
\mu(2l^{\rho}+q^{\rho})}{
\left[(l^0+\mu)^2-\boldsymbol{l}^2-m^2\right]
\left[(l^0+q^0+\mu)^2-(\boldsymbol{l}+\boldsymbol{q})^2-m^2\right]}
\nonumber\\&\hspace{0.3cm}
-4e^{2}\int\frac{d^Dl}{(2\pi)^{D}}\frac{2\mu^2g_{~0}^{\lambda}g_{~0}^{\tau}+(m^2-\mu^2)g^{\lambda\tau}}{
\left[(l^0+\mu)^2-\boldsymbol{l}^2-m^2\right]
\left[(l^0+q^0+\mu)^2-(\boldsymbol{l}+\boldsymbol{q})^2-m^2\right]}.
\end{align}
In terms of two generic one-loop tensor Feynman integrals $\mathscr{B}^{\rho}(q;m,\mu;m,\mu;\beta)$ and $\mathscr{B}^{\rho\sigma}(q;m,\mu;m,\mu;\beta)$, and a generic one-loop scalar Feynman integral $\mathcal{B}_{0}(q;m,\mu;m,\mu;\beta)$, the one-loop photon self-energy can be recast as
\begin{align}
&i\Pi^{\lambda\tau}(q;m,\mu;m,\mu;\beta)
\nonumber\\&
=\frac{-4ie^{2}}{(4\pi)^{D/2}}
\left\{\Big[g_{~\rho}^{\lambda}g_{~\sigma}^{\tau}-g^{\lambda\tau}g_{\rho\sigma}+g_{~\sigma}^{\lambda}g_{~\rho}^{\tau}\Big]
\Big[\mathscr{B}^{\rho\sigma}(q;m,\mu;m,\mu;\beta)+\mathscr{B}^{\rho}(q;m,\mu;m,\mu;\beta)q^\sigma\Big]
\right.\nonumber\\&\hspace{2cm}
+\mu\Big[g_{~\rho}^{\lambda}g_{~0}^{\tau}-g^{\lambda\tau}g_{\rho0}+g_{~0}^{\lambda}g_{~\rho}^{\tau}\Big]
\Big[2\mathscr{B}^{\rho}(q;m,\mu;m,\mu;\beta)+q^\rho \mathcal{B}_{0}(q;m,\mu;m,\mu;\beta)\Big]
\nonumber\\&\hspace{2cm}\left.
+\Big[2\mu^2g_{~0}^{\lambda}g_{~0}^{\tau}+(m^2-\mu^2)g^{\lambda\tau}\Big]\mathcal{B}_{0}(q;m,\mu;m,\mu;\beta)\right\}.
\label{ALLPSE}
\end{align}

After inserting $\mathscr{B}^{\rho}(p_{1};m_{1},\mu_{1};m_{2},\mu_{2};\beta)$ and $\mathscr{B}^{\rho\sigma}(p_{1};m_{1},\mu_{1};m_{2},\mu_{2};\beta)$ in Sec. \ref{ROLTFI} into Eq. (\ref{ALLPSE}) by setting $\mu_1=\mu_2=\mu$, $m_1=m_2=m$, and $p_{1}=q$, this one-loop photon self-energy can be straightforwardly performed by employing the \emph{generalized} Passarino-Veltman reduction. To put it differently, when $\mathscr{B}^{0}(q;m,\mu;m,\mu;\beta)$, $\mathscr{B}^{00}(q;m,\mu;m,\mu;\beta)$, $\mathcal{B}_{0}(q;m,\mu;m,\mu;\beta)$, and $\mathcal{A}_{0}(0;m,\mu;\beta)$ were previously obtained, $i\Pi^{\lambda\tau}(q;m,\mu;m,\mu;\beta)$ can be automatically assembled via the \emph{generalized} Passarino-Veltman reduction.

In order to further verify the \emph{generalized} Passarino-Veltman reduction, we check the Ward identity of one-loop photon self-energy at finite temperature and/or finite density. For convenience, we write down four components of $i\Pi^{\lambda\tau}(q;m,\mu;m,\mu;\beta)$, namely, 
\begin{align}
&i\Pi^{00}(q;m,\mu;m,\mu;\beta)
\nonumber\\&
=\frac{-4ie^{2}}{(4\pi)^{D/2}}
\left\{\Big.\mathscr{B}^{00}(q;m,\mu;m,\mu;\beta)+(2\mu+q^0)\mathscr{B}^{0}(q;m,\mu;m,\mu;\beta)
\right.\nonumber\\&\hspace{1.8cm}\left.
+\Big[\mu(\mu+q^0)+m^{2}\right]\mathcal{B}_{0}(q;m,\mu;m,\mu;\beta)
-g_{ab}\left[\mathscr{B}^{ab}(q;m,\mu;m,\mu;\beta)
+\mathscr{B}^{a}(q;m,\mu;m,\mu;\beta)q^{b}\Big]
\right\},\label{PSEF00}\\
&i\Pi^{i0}(q;m,\mu;m,\mu;\beta)
\nonumber\\&
=\frac{-4ie^{2}}{(4\pi)^{D/2}}
\left\{\Big.\mathscr{B}^{0i}(q;m,\mu;m,\mu;\beta)+\mathscr{B}^{i0}(q;m,\mu;m,\mu;\beta)+\mathscr{B}^{0}(q;m,\mu;m,\mu;\beta)q^i
\right.\nonumber\\&\hspace{1.8cm}\left.
+(2\mu+q^0)\mathscr{B}^{i}(q;m,\mu;m,\mu;\beta)
+\mu q^i\mathcal{B}_{0}(q;m,\mu;m,\mu;\beta)\Big.\right\}
\nonumber\\&
=i\Pi^{0i}(q;m,\mu;m,\mu;\beta),\label{PSEF0i}\\
&i\Pi^{ij}(q;m,\mu;m,\mu;\beta)
\nonumber\\&
=\frac{-4ie^{2}}{(4\pi)^{D/2}}
\left\{(g_{~a}^{i}g_{~b}^{j}-g^{ij}g_{ab}+g_{~b}^{i}g_{~a}^{j})
\Big[\mathscr{B}^{ab}(q;m,\mu;m,\mu;\beta)+\mathscr{B}^{a}(q;m,\mu;m,\mu;\beta)q^{b}\Big]
\right.\nonumber\\&\hspace{1.8cm}\left.
-g^{ij}\Big[\mathscr{B}^{00}(q;m,\mu;m,\mu;\beta)+(2\mu+q^0)\mathscr{B}^{0}(q;m,\mu;m,\mu;\beta)
+\left(\mu(\mu+q^0)-m^2\right)\mathcal{B}_{0}(q;m,\mu;m,\mu;\beta)\Big]\right\}.\label{PSEFij}
\end{align}

Obviously, $i\Pi^{00}(q;m,\mu;m,\mu;\beta)$, $i\Pi^{i0}(q;m,\mu;m,\mu;\beta)$, $i\Pi^{0i}(q;m,\mu;m,\mu;\beta)$, and $i\Pi^{ij}(q;m,\mu;m,\mu;\beta)$ can be expressed 
in terms of $\mathscr{B}^{0}(q;m,\mu;m,\mu;\beta)$ and $\mathscr{B}^{00}(q;m,\mu;m,\mu;\beta)$, $\mathcal{B}_{0}(q;m,\mu;m,\mu;\beta)$, and $\mathcal{A}_{0}(0;m,\mu;\beta)$. Substituting these results into $q_{\rho}\Pi^{\rho0}(q;m,\mu;m,\mu;\beta)$ and $q_{\rho}\Pi^{\rho j}(q;m,\mu;m,\mu;\beta)$, we have 
\begin{align}
q_{\rho}\Pi^{\rho0}(q;m,\mu;m,\mu;\beta)&=q_{0}\Pi^{00}(q;m,\mu;m,\mu;\beta)+q_{i}\Pi^{i0}(q;m,\mu;m,\mu;\beta)
\nonumber\\&
=q^{0}\Pi^{00}(q;m,\mu;m,\mu;\beta)-q^{i}\Pi^{i0}(q;m,\mu;m,\mu;\beta)=0,\\
q_{\rho}\Pi^{\rho j}(q;m,\mu;m,\mu;\beta)&=q_{0}\Pi^{0j}(q;m,\mu;m,\mu;\beta)+q_{i}\Pi^{ij}(q;m,\mu;m,\mu;\beta)
\nonumber\\&
=q^{0}\Pi^{0j}(q;m,\mu;m,\mu;\beta)-q^{i}\Pi^{ij}(q;m,\mu;m,\mu;\beta)=0.
\end{align}

These two relations can be further written in a compacted form as $q_{\rho}\Pi^{\rho\sigma}(q;m,\mu;m,\mu;\beta)=0$, which is nothing but the Ward identity. In this sense, we verified the \emph{generalized} Passarino-Veltman reduction by checking the Ward identity of one-loop photon self-energy at finite temperature and/or finite density.

It is subtle that besides the one-loop scalar Feynman integrals $\mathcal{A}_{0}(0;m,\mu;\beta)$ and $\mathcal{B}_{0}(q;m,\mu;m,\mu;\beta)$ that are adequate for expanding the one-loop pseudoscalar polarization function in the NJL model, the purely temporal components of generic one-loop tensor Feynman integrals $\mathscr{B}^{0}(q;m,\mu;m,\mu;\beta)$ and $\mathscr{B}^{00}(q;m,\mu;m,\mu;\beta)$ are also needed to form a complete set of basis in the one-loop photon self-energy in the $D$-dimensional QED. This difference originates from that the interaction term is $- \bar{\psi}\gamma^\rho \psi A_{\rho}$ in the QED (vector-type interaction) while $(\bar{\psi}\lambda^{a}\psi)^{2}+(\bar{\psi}i\gamma^{5}\lambda^{a}\psi)^{2}$ in the NJL model (scalar-type and pseudoscalar-type interactions). Roughly, in the relativistic QFTs with scalar-type or pseudoscalar-type interaction, the generic one-loop scalar Feynman integrals form a complete set of one-loop Feynman integrals, such as in the NJL model. By contrast, in the theories with vector-type or tensor-type interaction, the temporal components of generic one-loop tensor Feynman integrals are also needed, such as in the QED. Additionally, the \emph{generalized} Passarino-Veltman reduction presented here provides a procedure for automatic algebraic calculation, which in the Rehberg's scheme \cite{Rehberg1996PRC,Rehberg1996AOP}, the procedure can only be performed by hand. In this sense, the \emph{generalized} Passarino-Veltman reduction presented in Sec.\ref{ROLTFI} goes beyond the applicability of Rehberg's works \cite{Rehberg1996PRC,Rehberg1996AOP}.

We further comment on some aspects of practical application of employing the \emph{generalized} Passarino-Veltman reduction to the calculation of one-loop photon self-energy. In the context of QED for Weyl fermion (setting $m=0$ in the Lagrangian) in $D=4$ dimension, the author of the present paper explicitly calculated \cite{HRC2018} analytical expressions of one-loop photon self-energy $i\Pi^{\rho\sigma}(q;0,\mu;0,\mu;\beta=\infty)$ at zero temperature and finite density by utilizing the crude version of \emph{generalized} Passarino-Veltman reduction. The exact analytical expressions in Ref. \cite{HRC2018} automatically give rise to $i\Pi^{00}(q;0,\mu;0,\mu;\beta=\infty)$ \cite{HRCPlasmon2015}, $i\Pi^{ii}(q;0,\mu;0,\mu;\beta=\infty)$ [diagonal component of $i\Pi^{ij}(q;0,\mu;0,\mu;\beta=\infty)$] \cite{Agarwal2018}, $i\Pi^{0i}(q;0,\mu;0,\mu;\beta=\infty)$/$i\Pi^{i0}(q;0,\mu;0,\mu;\beta=\infty)$ \cite{Timm2019}, the parity-odd part of $i\Pi^{ij}(q;0,\mu;0,\mu;\beta=\infty)$ \cite{Kharzeev2009}, the parity-even part of $i\Pi^{ij}(q;0,\mu;0,\mu;\beta=\infty)$. Besides, under the hard dense loop approximation ($\mu\gg q^{0},|\boldsymbol{q}|$), it restores the results in Ref. \cite{DTSon2013}. All of these indicate that it is more efficient and general to calculate the one-loop photon energy once-for-all with the help of \emph{generalized} Passarino-Veltman reduction \cite{HRC2018} than by the conventional one-by-one method \cite{HRCPlasmon2015,Agarwal2018,Timm2019,Kharzeev2009,DTSon2013}.

\section{Summary and Discussions\label{Summary}}

In summary, we presented a \emph{generalized} Passarino-Veltman reduction for simplifying the generic one-loop tensor Feynman integrals in the one-loop Feynman diagrams of relativistic QFTs at finite temperature and/or finite density. It is explicitly demonstrated that the generic one-loop tensor Feynman integrals up to three-point can be decomposed as linear combinations of symmetric Lorentz-covariant tensor structures, non-Lorentz-covariant tensor structures, and their hybrid terms with the form factors being expressed by three generic one-loop scalar Feynman integrals $\mathcal{A}_{0}(0;m_1,\mu_1;\beta)$, $\mathcal{B}_{0}(p_{1};m_{1},\mu_{1};m_{2},\mu_{2};\beta)$, and $\mathcal{C}_{0}(p_1,p_2;m_1,\mu_1;m_2,\mu_2;m_3,\mu_3;\beta)$, and five generic one-loop tensor Feynman integrals $\mathscr{B}^{0}(p_{1};m_{1},\mu_{1};m_{2},\mu_{2};\beta)$, $\mathscr{B}^{00}(p_{1};m_{1},\mu_{1};m_{2},\mu_{2};\beta)$, $\mathscr{C}^{0}(p_1,p_2;m_1,\mu_1;m_2,\mu_2;m_3,\mu_3;\beta)$, $\mathscr{C}^{00}(p_1,p_2;m_1,\mu_1;m_2,\mu_2;m_3,\mu_3;\beta)$, and $\mathscr{C}^{000}(p_1,p_2;m_1,\mu_1;m_2,\mu_2;m_3,\mu_3;\beta)$. Since the generic one-loop scalar Feynman integrals, $\mathcal{A}_{0}(0;m_1,\mu_1;\beta)$, $\mathcal{B}_{0}(p_{1};m_{1},\mu_{1};m_{2},\mu_{2};\beta)$, and $\mathcal{C}_{0}(p_1,p_2;m_1,\mu_1;m_2,\mu_2;m_3,\mu_3;\beta)$, had already been analytically calculated in the Matsubara formalism \cite{Rehberg1996AOP}, after further analytically performing the purely temporal components of generic one-loop tensor Feynman integrals $\mathscr{B}^{0}(p_{1};m_{1},\mu_{1};m_{2},\mu_{2};\beta)$, $\mathscr{B}^{00}(p_{1};m_{1},\mu_{1};m_{2},\mu_{2};\beta)$, $\mathscr{C}^{0}(p_1,p_2;m_1,\mu_1;m_2,\mu_2;m_3,\mu_3;\beta)$, $\mathscr{C}^{00}(p_1,p_2;m_1,\mu_1;m_2,\mu_2;m_3,\mu_3;\beta)$, and $\mathscr{C}^{000}(p_1,p_2;m_1,\mu_1;m_2,\mu_2;m_3,\mu_3;\beta)$ in the Matsubara formalism, one can reduce all the one-loop Feynman diagrams up to three-point to these generic one-loop scalar Feynman integrals and the purely temporal components of generic one-loop tensor Feynman integrals. 

It is helpful to compare the \emph{generalized} Passarino-Veltman reduction above with its alternative representation. Following the essential spirit of \emph{conventional} Passarino-Veltman reduction that expresses the generic one-loop tensor Feynman integrals in terms of the Lorentz-covariant tensor structures dictated by the Lorentz symmetry \cite{CPVRS,Ellis,Denner2006}, one can reduce the spatial components of generic one-loop tensor Feynman integrals by utilizing the covariant tensor structures imposed by the $\mathrm{SO}(D-1)$ spatial rotation symmetry. In a spatial rotation covariant manner, the spatial components of generic one-loop tensor Feynman integrals $\mathscr{B}^{\rho}(p_{1};m_{1},\mu_{1};m_{2},\mu_{2};\beta)$ and $\mathscr{B}^{\rho\sigma}(p_{1};m_{1},\mu_{1};m_{2},\mu_{2};\beta)$ can be reduced on the same footing as

\begin{align}
\mathscr{B}^{i}(p_{1};m_{1},\mu_{1};m_{2},\mu_{2};\beta)&=\mathcal{B}_{1}(p_{1};m_{1},\mu_{1};m_{2},\mu_{2};\beta)p_{1}^{i},
\nonumber\\
\mathscr{B}^{0i}(p_{1};m_{1},\mu_{1};m_{2},\mu_{2};\beta)&=\mathscr{B}^{i0}(p_{1};m_{1},\mu_{1};m_{2},\mu_{2};\beta)
\nonumber\\&
=p_{1}^{0}p_{1}^{i}\mathcal{B}_{11}(p_{1};m_{1},\mu_{1};m_{2},\mu_{2};\beta)
+p_{1}^{i}\mathcal{B}_{12}(p_{1};m_{1},\mu_{1};m_{2},\mu_{2};\beta),
\nonumber\\
\mathscr{B}^{ij}(p_{1};m_{1},\mu_{1};m_{2},\mu_{2};\beta)&=-\delta^{ij}\mathcal{B}_{00}(p_{1};m_{1},\mu_{1};m_{2},\mu_{2};\beta)
+p_{1}^{i}p_{1}^{j}\mathcal{B}_{11}(p_{1};m_{1},\mu_{1};m_{2},\mu_{2};\beta).
\end{align}
Note that the other two components of the generic one-loop tensor Feynman integrals $\mathscr{B}^{\rho}(p_{1};m_{1},\mu_{1};m_{2},\mu_{2};\beta)$ and $\mathscr{B}^{\rho\sigma}(p_{1};m_{1},\mu_{1};m_{2},\mu_{2};\beta)$ are nothing but $\mathscr{B}^{0}(p_{1};m_{1},\mu_{1};m_{2},\mu_{2};\beta)$ and $\mathscr{B}^{00}(p_{1};m_{1},\mu_{1};m_{2},\mu_{2};\beta)$, which are related to the form factors $\mathcal{B}_{1}(p_{1};m_{1},\mu_{1};m_{2},\mu_{2})$, $\mathcal{B}_{2}(p_{1};m_{1},\mu_{1};m_{2},\mu_{2};\beta)$, $\mathcal{B}_{00}(p_{1};m_{1},\mu_{1};m_{2},\mu_{2};\beta)$, $\mathcal{B}_{11}(p_{1};m_{1},\mu_{1};m_{2},\mu_{2};\beta)$, $\mathcal{B}_{12}(p_{1};m_{1},\mu_{1};m_{2},\mu_{2};\beta)$, and $\mathcal{B}_{22}(p_{1};m_{1},\mu_{1};m_{2},\mu_{2};\beta)$ as 
\begin{align}
\mathscr{B}^{0}(p_{1};m_{1},\mu_{1};m_{2},\mu_{2};\beta)&=\mathcal{B}_{1}(p_{1};m_{1},\mu_{1};m_{2},\mu_{2};\beta)p_{1}^{0}
+\mathcal{B}_{2}(p_{1};m_{1},\mu_{1};m_{2},\mu_{2};\beta),
\nonumber\\
\mathscr{B}^{00}(p_{1};m_{1},\mu_{1};m_{2},\mu_{2};\beta)&=\mathcal{B}_{00}(p_{1};m_{1},\mu_{1};m_{2},\mu_{2};\beta)
+p_{1}^{0}p_{1}^{0}\mathcal{B}_{11}(p_{1};m_{1},\mu_{1};m_{2},\mu_{2};\beta)
\nonumber\\&\hspace{0.4cm}
+2p_{1}^{0}\mathcal{B}_{12}(p_{1};m_{1},\mu_{1};m_{2},\mu_{2};\beta)
+\mathcal{B}_{22}(p_{1};m_{1},\mu_{1};m_{2},\mu_{2};\beta).
\end{align}

Compared with the \emph{generalized} Passarino-Veltman reduction in Sec. \ref{ROLTFI}, which treats the spacetime components of generic one-loop tensor Feynman integrals on the same footing, the equivalent representation here restricts the reduction only to the spatial component of the generic one-loop tensor Feynman integrals, which is similar to the naive generalization of \emph{conventional} Passarino-Veltman reduction at finite temperature and/or finite density \cite{HRC2018} proposed by the author of the present paper.

\end{widetext}

The framework of \emph{generalized} Passarino-Veltman reduction for the generic one-loop tensor Feynman integrals can be straightforwardly extended to $N$-point, and hence can efficiently evaluate a huge amount of one-loop Feynman diagrams in physical systems described by the renormalizable relativistic QFTs at finite temperature and/or finite density, such as hot and dense quark matter \cite{QGP1,QGP2}. Based on the \emph{generalized} Passarino-Veltman reduction in this work, computer program packages can be developed for automatic algebraic calculation as that in the \emph{conventional} Passarino-Veltman reduction for the relativistic QFTs \cite{Oldenborgh1990,FeynArts1990,FeynCalc1991,LoopTools1999,QCDLoop2008,PackageX2015} or that in the generalization of nonrelativistic effective field theories at zero temperature and zero density \cite{Shtabovenko}.

It is emphasized that both the \emph{conventional} Passarino-Veltman reduction and the \emph{generalized} Passarino-Veltman reduction are based on continuous spacetime symmetry of the system. For the \emph{conventional} Passarino-Veltman reduction, it is the $\mathrm{SO}(1,D-1)$ (proper normal) Lorentz symmetry. While for the \emph{generalized} Passarino-Veltman reduction, it is the $\mathrm{SO}(D-1)$ spatial rotation symmetry broken down from Lorentz symmetry. If the $\mathrm{SO}(D-1)$ symmetry further breaks down to the $\mathrm{SO}(D-2)$ symmetry, then one can introduce another extra $D$-dimensional constant vector $v^{\rho}=(0;0,\cdots,0,1)$. Following a similar procedure, one can then reduce the generic one-loop tensor Feynman integrals after the symmetry-breaking of $\mathrm{SO}(D-1)$ spatial rotation.
Furthermore, the specific value of the dimension of spacetime $D$ does not affect the \emph{generalized} Passarino-Veltman reduction in this work, for example, it is valid for two distinct dimensions of physical interest, $D=3$ or $D=4$. However, if the dimension of spacetime is $D=2$ or the continuous spacetime symmetry of a physical system is less than $\mathrm{SO}(2)$ in $D$ dimension, there is no advantage of applying \emph{conventional} Passarino-Veltman reduction or \emph{generalized} Passarino-Veltman reduction to simplify the generic one-loop tensor Feynman integrals. Specifically, for the relativistic QFTs at zero temperature and zero density, the Lorentz covariance is restored for the absence of the preferred rest reference frame, and hence the $D$-dimensional constant vector always vanishes. As a consequence, the \emph{generalized} Passarino-Veltman reduction goes back to the \emph{conventional} Passarino-Veltman reduction.

This work opens up a new realm for the reductions of Feynman diagrams at loop level in the physical systems without Lorentz covariance. The \emph{generalized} Passarino-Veltman reduction presented in this work can be generalized to the physical systems of condensed matter described by pseudorelativistic QFTs at finite temperature and/or finite density, such as graphene \cite{Graphene} and silicene \cite{Silicene} in two spatial dimension and Dirac/Weyl semimetals \cite{TSM} in three spatial dimension. In addition, the generalizations include possible extensions  for one-loop tensor Feynman integrals in non-relativistic effective field theories and two-loop tensor Feynman integrals in relativistic QFTs. It is interesting to note that the reduction of one-loop tensor Feynman integrals in nonrelativistic effective field theories at zero temperature and zero density and the corresponding software toolkit \emph{FeynOnium} building upon \emph{FeynCalc} for automatic calculations  \cite{Shtabovenko} could be generalized to their counterparts at finite temperature and/or finite density by applying the essential spirits in this work. However, these important problems are beyond the focus of this work and deserve further study in the future.

\section*{Acknowledgments}

The author is grateful to Hong Guo, Sangyong Jeon, Dao-Neng Gao, Hongbao Zhang, Long Liang, and Yuanpei Lan for helpful discussions. This work is partially supported by the National Natural Science Foundation of China under Grant No. 11547200 and the China Scholarship Council under Grant No. 201608515061. The author would like to dedicate this paper to the memory of his past mother Xiu-Lan Xu.

\appendix
\allowdisplaybreaks[4]
\begin{widetext}

\section{EXPRESSIONS OF $\mathcal{F}_{1}(p_1;m_1,\mu_1;m_2,\mu_2;\beta)$ AND $\mathcal{F}_{2}(p_1;m_1,\mu_1;m_2,\mu_2;\beta)$ \label{App1}}

The numerator in the integrand of $\mathcal{F}_{1}(p_1;m_1,\mu_1;m_2,\mu_2;\beta)$ in Eq.(\ref{DefF1}) can be decomposed as
\begin{align}
p_1\cdot l&=p_1^0 l^0-\boldsymbol{l}\cdot\boldsymbol{p}_1
=p_1^0 l^0+\frac{\boldsymbol{l}^2+\boldsymbol{p}_1^2-(\boldsymbol{l}+\boldsymbol{p}_1)^2}{2}
\nonumber\\&
=\frac{\left(\mu_1^2-m_1^2\right)-\left[(\mu_2+p_1^0)^2-m_2^2\right]+\boldsymbol{p}_1^2}{2}
-\left(\mu_2-\mu_1\right)l^0+\frac{\mathcal{P}(l,p_1;m_2,\mu_2)-\mathcal{P}(l,0;m_1,\mu_1)}{2}.
\end{align}
Substituting this decomposition into the definition of $\mathcal{F}_{1}(p_1;m_1,\mu_1;m_2,\mu_2;\beta)$ in Eq.(\ref{DefF1}), we can directly express $\mathcal{F}_{1}(p_1;m_1,\mu_1;m_2,\mu_2;\beta)$ as
\begin{align}
&\mathcal{F}_1(p_1;m_1,\mu_1;m_2,\mu_2;\beta)
\nonumber\\&
=\frac{\left(\mu_1^2-m_1^2\right)-\left[(\mu_2+p_1^0)^2-m_2^2\right]+\boldsymbol{p}_1^2}{2}\mathcal{B}_{0}(p_1;m_1,\mu_1;m_2,\mu_2;\beta)
\nonumber\\&\hspace{0.4cm}
-\left(\mu_2-\mu_1\right)\mathscr{B}^{0}(p_1;m_1,\mu_1;m_2,\mu_2;\beta)
+\frac{\mathcal{A}_{0}(0;m_1,\mu_1;\beta)-\mathcal{A}_{0}(0;m_2,\mu_2;\beta)}{2}.
\end{align}

Different from the procedure for expressing $\mathcal{F}_{1}(p_1;m_1,\mu_1;m_2,\mu_2;\beta)$, we can straightforwardly obtain $\mathcal{F}_{2}(p_1;m_1,\mu_1;m_2,\mu_2;\beta)$ by defining
\begin{align}
&\mathcal{F}_{2}(p_1;m_1,\mu_1;m_2,\mu_2;\beta)=u_{\rho}\mathscr{B}^{\rho}(p_1;m_1,\mu_1;m_2,\mu_2;\beta)
=\mathscr{B}^{0}(p_1;m_1,\mu_1;m_2,\mu_2;\beta).
\end{align}

At zero temperature and zero density, $\mathcal{F}_1(p_1;m_1,\mu_1;m_2,\mu_2;\beta)$ reduces to
\begin{align}
&\mathcal{\tilde{F}}_1(p_1;m_1,\mu_1=0;m_2,\mu_2=0;\beta=\infty)=p_{1\rho}\mathscr{\tilde{B}}^{\rho}(p_1;m_1,\mu_1=0;m_2,\mu_2=0;\beta=\infty)
\nonumber\\&
=\frac{m_2^2-m_1^{2}-p_1^2}{2}\mathcal{\tilde{B}}_{0}(p_1;m_1,\mu_1=0;m_2,\mu_2=0;\beta=\infty)
\nonumber\\&\hspace{0.4cm}
+\frac{\mathcal{\tilde{A}}_{0}(0;m_1,\mu_1=0;\beta=\infty)-\mathcal{\tilde{A}}_{0}(0;m_2,\mu_2=0;\beta=\infty)}{2},
\end{align}
which can be obtained by setting $\mu_1=\mu_2=0$ and replacing the symbols without tilde ($T>0$) by the counterparts with tilde ($T=0$). Because $u^{\rho}$ automatically vanishes for the relativistic QFTs at zero temperature and zero density, it is not necessary to calculate $\mathcal{\tilde{F}}_2(p_1;m_1,\mu_1=0;m_2,\mu_2=0;\beta=\infty)$, the counterpart of $\mathcal{F}_{2}(p_1;m_1,\mu_1;m_2,\mu_2;\beta)$ at zero temperature.

\section{EXPRESSIONS OF $\mathcal{F}_{00}(p_1;m_1,\mu_1;m_2,\mu_2;\beta)$, $\mathcal{F}_{11}(p_1;m_1,\mu_1;m_2,\mu_2;
\beta)$, $\mathcal{F}_{12}(p_1;m_1,\mu_1;m_2,\mu_2;\beta)$, AND $\mathcal{F}_{22}(p_1;m_1,\mu_1;m_2,\mu_2;\beta)$ \label{App2}}

We express the numerator in the integrand of $\mathcal{F}_{00}(p_1;m_1,\mu_1;m_2,\mu_2;\beta)$ in Eq.(\ref{DefF00}) as
\begin{align}
l^2&=(l^0)^2-\boldsymbol{l}^2=-\left(\mu_1^2-m_1^2\right)-2\mu_1l^0+\mathcal{P}(l,0;m_1,\mu_1).
\end{align}
Substituting this decomposition into the definition of $\mathcal{F}_{00}(p_1;m_1,\mu_1;m_2,\mu_2;\beta)$, we have
\begin{align}
&\mathcal{F}_{00}(p_1;m_1,\mu_1;m_2,\mu_2;\beta)=g_{\rho\sigma}\mathscr{B}^{\rho\sigma}(p_1;m_1,\mu_1;m_2,\mu_2;\beta)
\nonumber\\&
=-\left(\mu_1^2-m_1^2\right)\mathcal{B}_{0}(p_1;m_1,\mu_1;m_2,\mu_2;\beta)-2\mu_1\mathscr{B}^{0}(p_1;m_1,\mu_1;m_2,\mu_2;\beta)
+\mathcal{A}_{0}(0;m_2,\mu_2;\beta).
\end{align}

Similarly, we recast the numerator in the integrand of $\mathcal{F}_{11}(p_1;m_1,\mu_1;m_2,\mu_2;\beta)$ in Eq.(\ref{DefF11}) as
\begin{align}
(p_1\cdot l)^2&
=\frac{1}{4}\left[\mathcal{P}(l,p_1;m_2,\mu_2)-\mathcal{P}(l,0;m_1,\mu_1)\right]^2
\nonumber\\&\hspace{0.4cm}
+\frac{\left\{\left(\mu_1^2-m_1^2\right)-\left[(\mu_2+p_1^0)^2-m_2^2\right]+\boldsymbol{p}_1^2\right\}^2}{4}
+\left(\mu_2-\mu_1\right)^2l^0l^0
\nonumber\\&\hspace{0.4cm}
-\left(\mu_2-\mu_1\right)\left[\mathcal{P}(l,p_1;m_2,\mu_2)-\mathcal{P}(l,0;m_1,\mu_1)\right]l^0
\nonumber\\&\hspace{0.4cm}
+\frac{\left(\mu_1^2-m_1^2\right)-\left[(\mu_2+p_1^0)^2-m_2^2\right]+\boldsymbol{p}_1^2}{2}
\left[\mathcal{P}(l,p_1;m_2,\mu_2)-\mathcal{P}(l,0;m_1,\mu_1)\right]
\nonumber\\&\hspace{0.4cm}
-\left(\mu_2-\mu_1\right)\left\{\left(\mu_1^2-m_1^2\right)-\left[(\mu_2+p_1^0)^2-m_2^2\right]+\boldsymbol{p}_1^2\right\}l^0.
\label{AppDecomF11}
\end{align}
It is helpful to perform two integrations
\begin{align}
&\int\frac{d^{D}l}{i\pi^{D/2}}\frac{1}{4}\left[\frac{\mathcal{P}(l,p_1;m_2,\mu_2)}{\mathcal{P}(l,0;m_1,\mu_1)}-1\right]
\nonumber\\&
=\int\frac{d^{D}l}{i\pi^{D/2}}\frac{1}{4}\left[\frac{2\left(\mu_2+p_1^0-\mu_1\right)(l^0+\mu_1)+(\mu_2+p_1^0-\mu_1)^2
+\left(m_1^2-m_2^2-\boldsymbol{p}_1^2\right)-2\boldsymbol{l}\cdot\boldsymbol{p}_1}{\mathcal{P}(l,0;m_1,\mu_1)}\right]
\nonumber\\&
=\frac{\left((\mu_2+p_1^0-\mu_1\right)^2+\left(m_1^2-m_2^2-\boldsymbol{p}_1^2\right)}{4}
\mathcal{A}_{0}(0;m_1,\mu_1;\beta),\label{AppDecomF11First}
\end{align}
and
\begin{align}
&\int\frac{d^{D}l}{i\pi^{D/2}}\frac{1}{4}\left[\frac{\mathcal{P}(l,0;m_1,\mu_1)}{\mathcal{P}(l,p_1;m_2,\mu_2)}-1\right]
\nonumber\\&
=\int\frac{d^{D}l}{i\pi^{D/2}}\frac{1}{4}\left[\frac{-2\left(\mu_2+p_1^0-\mu_1\right)(l^0+\mu_2+p_1^0)+(\mu_2+p_1^0-\mu_1)^2
-\left(m_1^2-m_2^2-\boldsymbol{p}_1^2\right)+2\boldsymbol{l}\cdot\boldsymbol{p}_1}{\mathcal{P}(l,p_1;m_2,\mu_2)}\right]
\nonumber\\&
=\frac{\left(\mu_2+p_1^0-\mu_1\right)^2-\left(m_1^2-m_2^2-\boldsymbol{p}_1^2\right)-2\boldsymbol{p}_1^2}{4}
\mathcal{A}_{0}(p_1;m_2,\mu_2;\beta)
\nonumber\\&
=\frac{\left(\mu_2+p_1^0-\mu_1\right)^2-\left(m_1^2-m_2^2+\boldsymbol{p}_1^2\right)}{4}\mathcal{A}_{0}(0;m_2,\mu_2;\beta),
\label{AppDecomF11Second}
\end{align}
where we have applied the following four relations
\begin{align}
\int\frac{d^{D}l}{i\pi^{D/2}}\frac{(l^0+\mu_1)}{\mathcal{P}(l,0;m_1,\mu_1)}&=0,\\
\int\frac{d^{D}l}{i\pi^{D/2}}\frac{\boldsymbol{l}\cdot\boldsymbol{p}_{1}}{\mathcal{P}(l,0;m_1,\mu_1)}&=0,\\
\int\frac{d^{D}l}{i\pi^{D/2}}\frac{(l^0+\mu_2+p_1^0)}{\mathcal{P}(l,p_1;m_2,\mu_2)}&=0,\\
\int\frac{d^{D}l}{i\pi^{D/2}}\frac{\boldsymbol{l}\cdot\boldsymbol{p}_{1}}{\mathcal{P}(l,p_1;m_2,\mu_2)}&
=-\boldsymbol{p}_1^2\mathcal{A}_{0}(0;m_2,\mu_2;\beta).
\end{align}

Substituting the decomposition in Eq.(\ref{AppDecomF11}) and two integrations in Eqs.(\ref{AppDecomF11First}) and (\ref{AppDecomF11Second}) into the definition of $\mathcal{F}_{11}(p_1;m_1,\mu_1;m_2,\mu_2;\beta)$ in Eq.(\ref{DefF11}), we have
\begin{align}
&\mathcal{F}_{11}(p_1;m_1,\mu_1;m_2,\mu_2;\beta)=p_{1\rho}p_{1\sigma}\mathscr{B}^{\rho\sigma}(p_1;m_1,\mu_1;m_2,\mu_2;\beta)
\nonumber\\&
=\frac{\left\{\left(\mu_1^2-m_1^2\right)-\left[(\mu_2+p_1^0)^2-m_2^2\right]+\boldsymbol{p}_1^2\right\}^2}{4}
\mathcal{B}_{0}(p_1;m_1,\mu_1;m_2,\mu_2;\beta)
\nonumber\\&\hspace{0.4cm}
+\left(\mu_2-\mu_1\right)^2\mathscr{B}^{00}(p_1;m_1,\mu_1;m_2,\mu_2;\beta)
\nonumber\\&\hspace{0.4cm}
-\left(\mu_2-\mu_1\right)\left\{\left(\mu_1^2-m_1^2\right)
-\left[(\mu_2+p_1^0)^2-m_2^2\right]+\boldsymbol{p}_1^2\right\}\mathscr{B}^{0}(p_1;m_1,\mu_1;m_2,\mu_2;\beta)
\nonumber\\&\hspace{0.4cm}
-\left(\mu_2-\mu_1\right)\left[\mathscr{A}^0(0;m_1,\mu_1;\beta)-\mathscr{A}^0(p_1;m_2,\mu_2;\beta)\right]
\nonumber\\&\hspace{0.4cm}
+\frac{\left(\mu_1^2-m_1^2\right)-\left[(\mu_2+p_1^0)^2-m_2^2\right]+\boldsymbol{p}_1^2}{2}
\left[\mathcal{A}_{0}(0;m_1,\mu_1;\beta)-\mathcal{A}_{0}(p_1;m_2,\mu_2;\beta)\right]
\nonumber\\&\hspace{0.4cm}
+\frac{\left(\mu_2+p_1^0-\mu_1\right)^2+\left(m_1^2-m_2^2-\boldsymbol{p}_1^2\right)}{4}
\mathcal{A}_{0}(0;m_1,\mu_1;\beta)
\nonumber\\&\hspace{0.4cm}
+\frac{\left(\mu_2+p_1^0-\mu_1\right)^2-\left(m_1^2-m_2^2+\boldsymbol{p}_1^2\right)}{4}
\mathcal{A}_{0}(p_1;m_2,\mu_2;\beta).
\end{align}

With the help of Eq.(\ref{EqAFinal}), we have
\begin{align}
&\mathcal{F}_{11}(p_1;m_1,\mu_1;m_2,\mu_2;\beta)
\nonumber\\&
=\frac{\left\{\left(\mu_1^2-m_1^2\right)-\left[(\mu_2+p_1^0)^2-m_2^2\right]+\boldsymbol{p}_1^2\right\}^2}{4}
\mathcal{B}_{0}(p_1;m_1,\mu_1;m_2,\mu_2;\beta)
\nonumber\\&\hspace{0.4cm}
+\left(\mu_2-\mu_1\right)^2\mathscr{B}^{00}(p_1;m_1,\mu_1;m_2,\mu_2;\beta)
\nonumber\\&\hspace{0.4cm}
-\left(\mu_2-\mu_1\right)\left\{\left(\mu_1^2-m_1^2\right)
-\left[(\mu_2+p_1^0)^2-m_2^2\right]+\boldsymbol{p}_1^2\right\}
\mathscr{B}^{0}(p_1;m_1,\mu_1;m_2,\mu_2;\beta)
\nonumber\\&\hspace{0.4cm}
+\frac{\left(m_2^2-m_1^2-p_1^2\right)-\left(\mu_1-\mu_2\right)^2-2p_1^0\left(\mu_1+\mu_2\right)}{4}
\mathcal{A}_{0}(0;m_1,\mu_1;\beta)
\nonumber\\&\hspace{0.4cm}
+\frac{\left(m_1^2-m_2^2+3p_1^2\right)-\left(\mu_1-\mu_2\right)^2+2p_1^0\left(\mu_1+\mu_2\right)}{4}
\mathcal{A}_{0}(0;m_2,\mu_2;\beta).
\end{align}

The numerator in the integrand of $\mathcal{F}_{12}(p_1;m_1,\mu_1;m_2,\mu_2;\beta)$ in Eq.(\ref{DefF12}) can be written as
\begin{align}
l^0(p_1\cdot l)&=\frac{\left(\mu_1^2-m_1^2\right)
-\left[(\mu_2+p_1^0)^2-m_2^2\right]+\boldsymbol{p}_1^2}{2}l^0
\nonumber\\&\hspace{0.4cm}
-\left(\mu_2-\mu_1\right)l^0l^0
+\frac{\mathcal{P}(l,p_1;m_2,\mu_2)l^0-\mathcal{P}(l,0;m_1,\mu_1)l^0}{2}.
\end{align}
Substituting this decomposition into the definition of $\mathcal{F}_{12}(p_1;m_1,\mu_1;m_2,\mu_2;\beta)$ in Eq.(\ref{DefF12}), we have
\begin{align}
&\mathcal{F}_{12}(p_1;m_1,\mu_1;m_2,\mu_2;\beta)
=p_{1\rho}u_{\sigma}\mathscr{B}^{\rho\sigma}(p_1;m_1,\mu_1;m_2,\mu_2;\beta)
\nonumber\\&
=\frac{\left(\mu_1^2-m_1^2\right)-\left[(\mu_2+p_1^0)^2-m_2^2\right]+\boldsymbol{p}_1^2}{2}
\mathscr{B}^{0}(p_1;m_1,\mu_1;m_2,\mu_2;\beta)
\nonumber\\&\hspace{0.4cm}
-\left(\mu_2-\mu_1\right)\mathscr{B}^{00}(p_1;m_1,\mu_1;m_2,\mu_2;\beta)
+\frac{\mathscr{A}^0(0;m_1,\mu_1;\beta)-\mathscr{A}^0(p_1;m_2,\mu_2;\beta)}{2}
\nonumber\\&
=\frac{\left(\mu_1^2-m_1^2\right)-\left[(\mu_2+p_1^0)^2-m_2^2\right]+\boldsymbol{p}_1^2}{2}
\mathscr{B}^{0}(p_1;m_1,\mu_1;m_2,\mu_2;\beta)
\nonumber\\&\hspace{0.4cm}
-\left(\mu_2-\mu_1\right)\mathscr{B}^{00}(p_1;m_1,\mu_1;m_2,\mu_2;\beta)
\nonumber\\&\hspace{0.4cm}
+\frac{-\mu_1\mathcal{A}_0(0;m_1,\mu_1;\beta)+(\mu_2+p_{1}^{0})\mathcal{A}_0(0;m_2,\mu_2;\beta)}{2}.
\end{align}

Different from the above decomposition procedures, $\mathcal{F}_{22}(p_1;m_1,\mu_1;m_2,\mu_2;\beta)$ can be straightforwardly obtained by defining
\begin{align}
\mathcal{F}_{22}(p_1;m_1,\mu_1;m_2,\mu_2;\beta)&=u_{\rho}u_{\sigma}\mathscr{B}^{\rho\sigma}(p_1;m_1,\mu_1;m_2,\mu_2;\beta)
=\mathscr{B}^{00}(p_1;m_1,\mu_1;m_2,\mu_2;\beta).
\end{align}

At zero temperature and zero density, $\mathcal{F}_{00}(p_1;m_1,\mu_1;m_2,\mu_2;\beta)$ and $\mathcal{F}_{11}(p_1;m_1,\mu_1;m_2,\mu_2;\beta)$ can be rewritten as
\begin{align}
&\mathcal{\tilde{F}}_{00}(p_1;m_1,\mu_1=0;m_2,\mu_2=0;\beta=\infty)
\nonumber\\&
=m_1^2\mathcal{\tilde{B}}_{0}(p_1;m_1,\mu_1=0;m_2,\mu_2=0;\beta=\infty)+\mathcal{\tilde{A}}_{0}(0;m_2,\mu_2=0;\beta=\infty),\\
&\mathcal{\tilde{F}}_{11}(p_1;m_1,\mu_1=0;m_2,\mu_2=0;\beta=\infty)
\nonumber\\&
=\frac{\left(m_2^2-m_1^2-p_1^2\right)^2}{4}\mathcal{\tilde{B}}_{0}(p_1;m_1,\mu_1=0;m_2,\mu_2=0;\beta=\infty)
\nonumber\\&\hspace{0.4cm}
+\frac{m_2^2-m_1^2-p_1^2}{4}
\mathcal{A}_{0}(0;m_1,\mu_1=0;\beta=\infty)
+\frac{m_1^2-m_2^2+3p_1^2}{4}\mathcal{\tilde{A}}_{0}(0;m_2,\mu_2=0;\beta=\infty),
\end{align}
which can be obtained by setting $\mu_1=\mu_2=0$ and replacing the symbols without tilde ($T>0$) by the counterparts with tilde ($T=0$). Because $u^{\rho}$ and $u^{\sigma}$ automatically vanish for the relativistic QFTs at zero temperature and zero density, it is not necessary to calculate $\mathcal{\tilde{F}}_{12}(p_1;m_1,\mu_1=0;m_2,\mu_2=0;\beta=\infty)$ and $\mathcal{\tilde{F}}_{22}(p_1;m_1,\mu_1=0;m_2,\mu_2=0;\beta=\infty)$, the counterparts of $\mathcal{F}_{12}(p_1;m_1,\mu_1;m_2,\mu_2)$ and $\mathcal{F}_{22}(p_1;m_1,\mu_1;m_2,\mu_2)$ at zero temperature.

\section{EXPRESSIONS OF  $\mathcal{K}_1(p_1,p_2;m_1,\mu_1;m_2,\mu_2;m_3,\mu_3;\beta)$, $\mathcal{K}_2(p_1,p_2;m_1,\mu_1;m_2,\mu_2;m_3,\mu_3;\beta)$ AND $\mathcal{K}_3(p_1,p_2;m_1,\mu_1;m_2,\mu_2;m_3,\mu_3;\beta)$ \label{App3}}

One can express $\mathcal{K}_1(p_1,p_2;m_1,\mu_1;m_2,\mu_2;m_3,\mu_3;\beta)$, $\mathcal{K}_2(p_1,p_2;m_1,\mu_1;m_2,\mu_2;m_3,\mu_3;\beta)$, and $\mathcal{K}_3(p_1,p_2;m_1,\mu_1;m_2,\mu_2;m_3,\mu_3;\beta)$ in a parallel procedure. The numerator in the integrand of $\mathcal{K}_1(p_1,p_2;m_1,\mu_1;m_2,\mu_2;m_3,\mu_3;\beta)$ can be decomposed as
\begin{align}
p_1\cdot l&=p_1^0 l^0-2\boldsymbol{l}\cdot\boldsymbol{p}_1
=p_1^0 l^0+\frac{\boldsymbol{l}^2+\boldsymbol{p}_1^2-(\boldsymbol{l}+\boldsymbol{p}_1)^2}{2}
\nonumber\\&
=\frac{\left(\mu_1^2-m_1^2\right)-\left((\mu_2+p_1^0)^2-m_2^2\right)+\boldsymbol{p}_1^2}{2}
-\left(\mu_2-\mu_1\right)l^0
+\frac{\mathcal{P}(l,p_1;m_2,\mu_2)-\mathcal{P}(l,0;m_1,\mu_1)}{2}.
\end{align}
Substituting it into the definition of $\mathcal{K}_1(p_1,p_2;m_1,\mu_1;m_2,\mu_2;m_3,\mu_3;\beta)$ in Eq.(\ref{DefG1}), we obtain
\begin{align}
&\mathcal{K}_1(p_1,p_2;m_1,\mu_1;m_2,\mu_2;m_3,\mu_3;\beta)=p_{1\rho}\mathscr{C}^{\rho}(p_1,p_2;m_1,\mu_1;m_2,\mu_2;m_3,\mu_3;\beta)
\nonumber\\&
=\frac{\left(\mu_1^2-m_1^2\right)-\left((\mu_2+p_1^0)^2-m_2^2\right)+\boldsymbol{p}_1^2}{2}
\mathcal{C}_{0}(p_1,p_2;m_1,\mu_1;m_2,\mu_2;m_3,\mu_3;\beta)
\nonumber\\&\hspace{0.4cm}
-\left(\mu_2-\mu_1\right)\mathscr{C}^{0}(p_1,p_2;m_1,\mu_1;m_2,\mu_2;m_3,\mu_3;\beta)
\nonumber\\&\hspace{0.4cm}
+\frac{\mathcal{B}_{0}(p_{1}+p_{2};m_1,\mu_1;m_3,\mu_3;\beta)-\mathcal{B}_{0}(p_{2};m_2,\mu_2;m_3,\mu_3;\beta)}{2}.
\end{align}

Similarly, the numerator in the integrand of $\mathcal{K}_2(p_1,p_2;m_1,\mu_1;m_2,\mu_2;m_3,\mu_3;\beta)$ can be recast as
\begin{align}
p_{2}\cdot l&
=\frac{\left[(\mu_2+p_1^0)^2-m_2^2-\boldsymbol{p}_1^2\right]
-\left[(\mu_3+p_1^0+p_2^0)^2-m_3^2-(\boldsymbol{p}_1+\boldsymbol{p}_2)^2\right]}{2}
\nonumber\\&\hspace{0.4cm}
-\left(\mu_3-\mu_2\right)l^0+\frac{\mathcal{P}(l,p_1+p_2;m_{3},\mu_{3})-\mathcal{P}(l,p_1;m_{2},\mu_{2})}{2}.
\end{align}
Substituting it into the definition of $\mathcal{K}_2(p_1,p_2;m_1,\mu_1;m_2,\mu_2;m_3,\mu_3;\beta)$ in Eq.(\ref{DefG2}), we arrive at
\begin{align}
&\mathcal{K}_2(p_1,p_2;m_1,\mu_1;m_2,\mu_2;m_3,\mu_3;\beta)=p_{2\rho}\mathscr{C}^{\rho}(p_1,p_2;m_1,\mu_1;m_2,\mu_2;m_3,\mu_3;\beta)
\nonumber\\&
=\frac{\left[(\mu_2+p_1^0)^2-m_2^2-\boldsymbol{p}_1^2\right]
-\left[(\mu_3+p_1^0+p_2^0)^2-m_3^2-(\boldsymbol{p}_1+\boldsymbol{p}_2)^2\right]}{2}
\mathcal{C}_{0}(p_1,p_2;m_1,\mu_1;m_2,\mu_2;m_3,\mu_3;\beta)
\nonumber\\&\hspace{0.4cm}
-\left(\mu_3-\mu_2\right)\mathscr{C}^{0}(p_1,p_2;m_1,\mu_1;m_2,\mu_2;m_3,\mu_3;\beta)
\nonumber\\&\hspace{0.4cm}
+\frac{\mathcal{B}_{0}(p_{1};m_1,\mu_1;m_2,\mu_2;\beta)-\mathcal{B}_{0}(p_{1}+p_{2};m_1,\mu_1;m_3,\mu_3;\beta)}{2}.
\end{align}

Different from the above procedures, $\mathcal{K}_{3}(p_1,p_2;m_1,\mu_1;m_2,\mu_2;m_3,\mu_3;\beta)$ can be straightforwardly obtained by defining
\begin{align}
&\mathcal{K}_3(p_1,p_2;m_1,\mu_1;m_2,\mu_2;m_3,\mu_3;\beta)=u_{\rho}\mathscr{C}^{\rho}(p_1,p_2;m_1,\mu_1;m_2,\mu_2;m_3,\mu_3;\beta)
\nonumber\\
&=\mathscr{C}^{0}(p_1,p_2;m_1,\mu_1;m_2,\mu_2;m_3,\mu_3;\beta).
\end{align}

At zero temperature and zero density, $\mathcal{K}_1(p_1,p_2;m_1,\mu_1;m_2,\mu_2;m_3,\mu_3;\beta)$ and $\mathcal{K}_2(p_1,p_2;m_1,\mu_1;m_2,\mu_2;m_3,\mu_3;\beta)$ can be rewritten as
\begin{align}
&\mathcal{\tilde{K}}_1(p_1,p_2;m_1,\mu_1=0;m_2,\mu_2=0;m_3,\mu_3=0;\beta=\infty)
\nonumber\\&
=\frac{m_2^2-m_1^2-p_1^2}{2}\mathcal{\tilde{C}}_{0}(p_1,p_2;m_1,\mu_1=0;m_2,\mu_2=0;m_3,\mu_3=0;\beta=\infty)
\nonumber\\&\hspace{0.4cm}
+\frac{\mathcal{\tilde{B}}_{0}(p_{1}+p_{2};m_1,\mu_1=0;m_3,\mu_3=0;\beta=\infty)
-\mathcal{\tilde{B}}_{0}(p_{2};m_2,\mu_2=0;m_3,\mu_3=0;\beta=\infty)}{2},\\
&\mathcal{\tilde{K}}_2(p_1,p_2;m_1,\mu_1=0;m_2,\mu_2=0;m_3,\mu_3=0;\beta=\infty)
\nonumber\\&
=\frac{m_3^2-m_2^2+p_1^2-(p_1+p_2)^2}{2}\mathcal{C}_{0}(p_1,p_2;m_1,\mu_1=0;m_2,\mu_2=0;m_3,\mu_3=0;\beta=\infty)
\nonumber\\&\hspace{0.4cm}
+\frac{\mathcal{\tilde{B}}_{0}(p_{1};m_1,\mu_1=0;m_2,\mu_2=0;\beta=\infty)
-\mathcal{\tilde{B}}_{0}(p_{1}+p_{2};m_1,\mu_1=0;m_3,\mu_3=0;\beta=\infty)}{2},
\end{align}
which can be obtained by setting $\mu_1=\mu_2=\mu_3=0$ and replacing the symbols without tilde ($T>0$) by the counterparts with tilde ($T=0$). Because $u^{\rho}$ automatically vanishes for the relativistic QFTs at zero temperature and zero density, it is not necessary to calculate $\mathcal{\tilde{K}}_3(p_1,p_2;m_1,\mu_1=0;m_2,\mu_2=0;m_3,\mu_3=0;\beta=\infty)$, the counterpart of $\mathcal{K}_3(p_1,p_2;m_1,\mu_1;m_2,\mu_2;m_3,\mu_3;\beta)$ at zero temperature.

\end{widetext}


\end{CJK}

\end{document}